\documentclass[aps,pre,groupedaddress,showpacs,amsmath,eqsecnum]{revtex4}

\usepackage{amssymb}
\usepackage{amsmath}
\usepackage{graphicx}
\usepackage{amsbsy}
\usepackage{color}

\newcommand{\HS}{\text{HS}}

\newcommand{\SHS}{\text{SHS}}

\newcommand{\SPS}{\text{SPS}}

\newcommand{\SW}{\text{SW}}

\begin{document}
\title{Penetrable Square-Well fluids: Exact results in one dimension}
\author{Andr\'es Santos}
\email{andres@unex.es} \homepage{http://www.unex.es/fisteor/andres/}
\affiliation{Departamento de F\'isica, Universidad de Extremadura,
E-06071 Badajoz, Spain}
\author{Riccardo Fantoni}
\email{rfantoni@unive.it}
\affiliation{Dipartimento di Chimica Fisica, Universit\`a di Venezia,
Calle Larga S. Marta DD2137, I-30123 Venezia, Italy}
\author{Achille Giacometti}
\email{achille@unive.it}
\affiliation{Dipartimento di Chimica Fisica, Universit\`a di Venezia,
Calle Larga S. Marta DD2137, I-30123 Venezia, Italy}
\date{\today}

\begin{abstract}
We introduce a model of attractive penetrable spheres by adding
 {a} short range attractive square well outside a penetrable
core, and we provide a detailed analysis of structural and
thermodynamical properties in one dimension using the exact
 {impenetrable} counterpart as a starting point. The model is
expected to describe star polymers in regimes of good and moderate
solvent under dilute conditions. We derive the exact coefficients of
a low density expansion up to second order for the radial
distribution function and up to fourth order in the virial
expansion. These exact results are  {used as a benchmark to test
the reliability of  approximate theories (Percus-Yevick and
hypernetted chain)}. Notwithstanding the lack of an exact solution
 {for arbitrary densities}, our results are  expected to be
rather precise within a wide range of temperatures and densities. A
detailed analysis of  {some} limiting cases is carried out. In
particular we provide a complete solution of the sticky
penetrable-sphere model in one dimension up to the same order in
density. The issue of Ruelle's thermodynamics stability is analyzed
and the region of a well defined thermodynamic limit is identified.
\end{abstract}



\maketitle
\section{Introduction}
\label{sec:intro}
Unlike simple liquids, where two-body potentials describe
interactions at the atomistic levels, complex liquid interactions
are always a result of an average process over the microscopic
degrees of freedom. As a result, highly simplified models often
accurately describe a number of experimental features ranging from
structural to thermophysical properties. Examples include colloidal
dispersions, macromolecules, and combinations of the two
\cite{Barrat03}. A noteworthy feature of these system is that the
hard-core repulsive barrier for very short range is not an
essential ingredient of the model. In the case of highly ramified
polymers in good solvent ({star} polymers), for instance, the
center-of-mass of two polymer chains can  {be at a} distance much
smaller than their respective radii of gyration and  {they are} well
described by an effective Gaussian interaction
 \cite{Likos01}. The simplest of this class of minimal bounded
potential is the so-called penetrable spheres (PS) \cite{Marquest89}
that has attracted considerable attention in the last few years (see
e.g. Ref.\ \cite{Malijevsky06} and references therein). In this case
the infinite barrier of the hard sphere (HS) potential is replaced
by a finite one, thus allowing for a finite probability of
penetrating inside the core.

A major advantage of  {the PS} potential is, of course,
simplicity. On the other hand, it lacks  an attractive part which is
expected to be relevant in such a complex environment in view of the
ubiquity of van der Waals dispersion forces. The purpose of the
present work is to address this point by proposing a variation of
the PS model in which a square well (SW) is added outside the core.
This model, hereafter referred to as penetrable square-well (PSW),
has an extremely rich phenomenology notwithstanding its simplicity,
including a number of interesting limiting cases as will be
discussed later on.

One-dimensional  {bounded} interactions do not belong to the
class of nearest-neighbor fluids for which the total potential
energy can be written as
\begin{eqnarray}
U_N(x_1,\ldots,x_N)=\sum_{i=1}^{N-1}\phi(|x_{i+1}-x_i|)~,
\label{introduction:eq1}
\end{eqnarray}
where $\phi(r)$ is the pair potential and $\{x_i, i=1,\ldots,N\}$
are the coordinates of the $N$ particles confined in a segment of
length $L$, which eventually may be let to go to infinity. A
necessary (but not sufficient) condition for a one-dimensional fluid
to  {satisfy Eq.\ \eqref{introduction:eq1}} is to be a hard-core
fluid, {i.e.}, a fluid made of particles which cannot penetrate one
another due to the existence of an infinite repulsive potential
barrier in $\phi(r)$.

Nearest-neighbor fluids admit an analytic exact
statistical-mechanical solution in one-dimension \cite{Salsburg53}:
the partition function, equation of state, and correlation functions
of any order can be calculated analytically from the knowledge of
the pair potential. In addition to HS
\cite{Salsburg53,Tonks36,Feynman72,Corti98,Lebowitz71,Santos07a},
both structural and thermophysical properties can be analytically
obtained in one dimension for Baxter's sticky hard-sphere
(SHS) potential \cite{Seaton86,Yuste93}, and for the SW potential
\cite{Heying04}, but the technique permits in principle the analysis
of a large class of nearest-neighbor one-dimensional potentials.

In the absence of the nearest-neighborhood constraint  {(as
happens with bounded potentials)}, the situation is far more complex
and we are not aware of any general analytical approach to the
problem, even in one dimension. As a matter of fact there exist only
a few examples of  analytically solvable one-dimensional models of
this type, which include the Kac potential \cite{Lieb66} and the
Coulomb potential \cite{Fantoni03,note1}.  For PS, it was observed
\cite{Malijevsky06,Santos07} that the exact analytic solution for HS
can be efficiently exploited to build a rather precise, albeit
approximate, solution of the penetrable counterpart. This analysis
is here extended to PSW interactions. Using a low density expansion
and the corresponding exact solution for the SW problem, we derive
the exact result up to the second order in a density expansion of
the radial distribution function and up to fourth order in the
virial expansion  {of the equation of state}.  {These exact
low density calculations} are contrasted with approximate theories
such as the Percus-Yevick (PY) and the hypernetted chain (HNC)
closures, thus providing an assessment on the relative reliability
of both  approximations and the low density expansion. As a
preliminary simplified step to our calculation, we also examine the
penetrable counterpart of the SHS problem, denoted as sticky
penetrable spheres (SPS) in the following, which provides a
guideline to tackle the more difficult PSW problem.

The introduction of an attractive part of the potential into a
penetrable interaction raises the important issue of  {the
existence of} a well defined thermodynamic limit
\cite{Fisher66,Ruelle69}. We address this problem for the PSW model
and provide compelling arguments to identify the stability region,
which is guaranteed for a sufficiently small ($\approx 0.5$) ratio
between the attractive and repulsive energy scales and arbitrary
values of the other parameters.

The remaining of the paper is structured as follows. In Section
\ref{sec:psw} we introduce the model along with all its limiting
cases (including the SPS fluid) and we study its stability. Section
\ref{sec:formalism} briefly accounts for the main equations
necessary for the analytical solution of the nearest-neighbor class
of fluids with arbitrary interactions. The exact solution of the
one-dimensional SHS potential is derived within this general
approach in Section \ref{sec:exactshs} and this is used to obtain
the corresponding low density solution of the SPS in Section
\ref{sec:exactsps}. A similar analysis is carried out in Section
\ref{sec:exactpsw} for PSW and the results are contrasted
with those stemming from PY and HNC closures. Section
\ref{sec:conclusions} contains some closing remarks, whereas some of
the more technical details are confined in suitable appendices.

\section{The penetrable square-well (PSW) model}
\label{sec:psw}
\begin{figure}
\includegraphics[width=.4 \columnwidth]{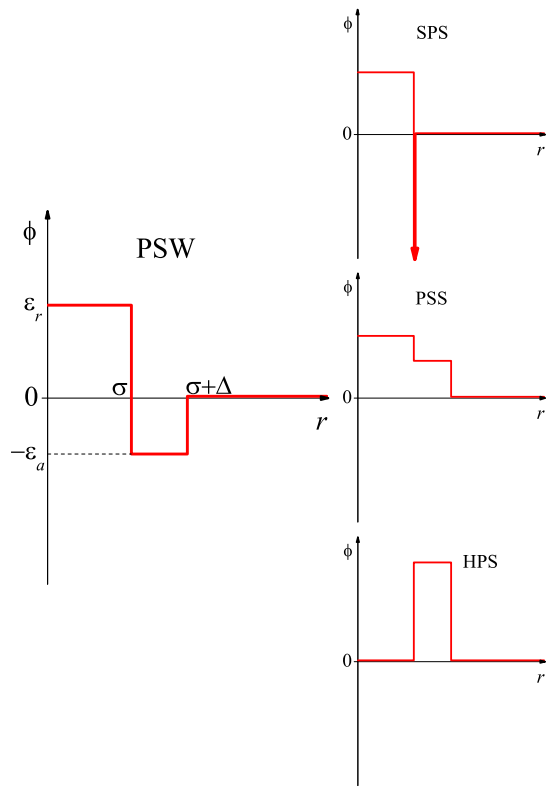}
\caption{(Color online) Sketch of the  penetrable square-well (PSW) 
potential (left column). The right column shows a few limiting cases: the 
sticky penetrable spheres (SPS)
potential ($\epsilon_a\to\infty$ and $\Delta\to 0$), the penetrable square-shoulder (PSS)
potential ($\epsilon_r>-\epsilon_a>0$), and the  hollow hard-spheres (HPS) potential
($\epsilon_r=0$ and $\epsilon_a <0$).
\label{potential}}
\end{figure}

The penetrable square-well (PSW) model is defined by the following
pair potential (see Fig.\ \ref{potential})
\begin{eqnarray}
\phi\left(r\right)=\left\{
\begin{array}{ll}
+\epsilon_r, & r<\sigma,\\
-\epsilon_a, & \sigma<r<\sigma+\Delta,\\
0  ,         & r>\sigma+\Delta,
\end{array}\right.
\label{psw:eq1}
\end{eqnarray}
where $\epsilon_r$ and $\epsilon_a$ are two positive constants
accounting for the repulsive and attractive parts of the potential,
respectively. Here $\sigma$ is the  {diameter of the sphere (length
of the rod  in one dimension)} and $\Delta < \sigma$ is the width of
the well. This model has a number of relevant limiting cases. When
$\epsilon_r \to \infty$ it reduces to a square-well (SW) fluid,
whereas $\epsilon_a \to 0$ yields the  {penetrable-sphere (PS) model
studied in Ref.\ \cite{Malijevsky06} in the one-dimensional case}.
In addition it gives rise to an interesting variation, referred to
as ``sticky  penetrable spheres'' (SPS), within an
appropriate limit of infinite well  {depth} of vanishing width (see
below). Finally, we recover the
 {hard-sphere (HS)} fluid in the combined limit $\epsilon_r \to
\infty$ and $\epsilon_a \to 0$.

It is worthwile noting that the PSW model (and its variants) considered here
is different from other apparently similar models like the Widom-Rowlinson's
of interpenetrating spheres \cite{Widom70}, the concentric-shell model
\cite{Torquato84}, or the permeable-sphere model \cite{Rikvold85},

As usual, a very important role is played by the Mayer function
\begin{eqnarray}
f\left(r\right) &=& e^{- \beta \phi\left(r\right) } -1,
\label{psw:eq2}
\end{eqnarray}
where $\beta =1/k_B T$ is the inverse of the thermal energy ($k_B$
is the Boltzmann constant and $T$ is the absolute temperature). In
the present model, this becomes
\begin{eqnarray} \nonumber
f\left(r\right)&=&\left\{
\begin{array}{ll}
-\gamma_r, & r<\sigma,\\
+\gamma_a, & \sigma<r<\sigma+\Delta,\\
0 ,          & r>\sigma+\Delta,
\end{array}\right .\\ \label{psw:eq3}
&=& \gamma_r
f_{\HS}(r)+\gamma_a[\Theta(r-\sigma)-\Theta(r-\sigma-\Delta)]~,
\end{eqnarray}
 {where}
\begin{equation}
\gamma_r= 1-e^{-\beta\epsilon_r} \label{gamma_r}
\end{equation}
is the parameter measuring the degree of penetrability varying
between 0 (free penetrability) and 1 (impenetrability) and
\begin{equation}
\gamma_a= e^{\beta\epsilon_a}-1
\label{gamma_a}
\end{equation}
plays a similar role for the attractive part. Here
$f_{\HS}(r)=\Theta(r-\sigma)-1$ is the Mayer function for a
 {HS} model, which can then be recovered in the limit $\gamma_r
\to 1$ and $\gamma_a \to 0$, and $\Theta(r)$ is the usual step
function equal to $1$ for $r>0$ and $0$ otherwise. It also proves
convenient to introduce the ratio $\gamma = \gamma_a/\gamma_r$ which
is a measure of the depth of the attractive well,  {relative to
the ``penetrability'' of the core. In that way, Eq.\ \eqref{psw:eq3}
can be rewritten as}
\begin{eqnarray}
f(r)&=&\gamma_r\left\{f_{\HS}(r)+\gamma[\Theta(r-\sigma)-\Theta(r-\sigma-\Delta)]\right\}\nonumber\\
&=&\gamma_r f_{\SW}(r),
\label{f_psw}
\end{eqnarray}
 {where $f_{\SW}(r)$ is the Mayer function of a SW fluid with
the change $\gamma_a\to\gamma$.}

 {Although the PSW model can be defined for any dimensionality
of the system, throughout the remainder of this paper we will
specialize to the one-dimensional case.}
\subsection{The thermodynamic stability issue}
\label{subsec:stability}
As anticipated in the Introduction, in our model we need to make
sure that the system is always \textit{stable} in the sense that the
 {total} energy is always bounded from below by $-N B$, $N$
being the number of particles and $B$ being an arbitrary positive
constant \cite{Ruelle69}. The physical origin of such an instability
can be traced back to the fact that a soft core allows the
possibility of a ``collapsed-state'' where the energy is no longer
proportional to the number of particles $N$ and a well-defined
thermodynamic limit may not exist. In a classic paper, Fisher and
Ruelle \cite{Fisher66} provided a set of conditions on the pair
potentials which are sufficient for stability, but the actual
implementation of such conditions in soft-core systems is far from
being trivial as was recently shown for Gaussian-core models
\cite{Louis00} and Lennard-Jones fluids \cite{Heyes07}.

In the PSW model, the issue is clearly related to the interplay
between the two energy scales, $\epsilon_r$ and $\epsilon_a$, for
the repulsive and attractive parts of the potential. As shown in
Appendix \ref{app:ruelle}, we predict that the system \textit{might
be unstable} when $\epsilon_r < 2 \epsilon_a$ whereas we prove that
it is \textit{certainly stable} in the opposite case $\epsilon_r > 2
\epsilon_a$.
\subsection{The sticky limit: the SPS model}
\label{subsec:sps}
It is instructive at this point to consider a particular limit of
the PSW model which will be referred to as the sticky
penetrable-sphere (SPS) model  {(see Fig.\ \ref{potential})}. This
is a variation of a widely used \textit{sticky hard-sphere} (SHS)
model introduced long time ago by Baxter \cite{Baxter68}, which has
proven to be extremely useful in the framework of complex
fluids, recently even in its anisotropic version
\cite{Fantoni07}. The simplest way of introducing it is at the level
of the Mayer function [see Eq.\ (\ref{psw:eq3})], which becomes
\begin{eqnarray}
f_{\SHS}\left(r\right)&=& f_{\HS}\left(r\right)+\alpha \sigma
\delta_+ \left(r-\sigma\right)~
\label{sps:eq1}
\end{eqnarray}
where
\begin{eqnarray}
\delta_+\left(r\right) &=&\lim_{a \to 0^+}\frac{\Theta(r)-
\Theta(r-a)}{a},
\label{sps:eq2}
\end{eqnarray}
The relation with the SPS model is then provided by
\begin{eqnarray}
f_{\SPS}\left(r\right) &=& \gamma_r f_{\SHS}\left(r\right)
\label{sps:eq3}
\end{eqnarray}
In the original SHS model \cite{Baxter68}, $\alpha=1/12 \tau>0$,
($\tau$ playing the role of an effective temperature) but the
connection with the PSW model is readily achieved from Eq.\
 {(\ref{f_psw})} by considering the  {limits $\Delta \to 0$
and $\epsilon_r\to\infty$} so that
$\alpha=(\gamma_a/\gamma_r)(\Delta/\sigma)$ remains finite. In spite
of its usefulness, the SHS model is known to suffer from some
mathematical drawbacks, the most important of them being that it is
unstable in spatial dimensions greater than $1$, as pointed out by
Stell \cite{Stell91}, in view of the divergence of the virial
coefficient corresponding to a close packed configuration. For the
SPS model we will be able to achieve a number of exact results which
can be exploited as a guideline for the more complex PSW model.
\subsection{Other limiting cases}
\label{subsec:other}
\begin{table}
\caption{\label{tab:models} Summary of the models.}
\begin{ruledtabular}
\begin{tabular}{lcccc}
Model & Acronym &$\epsilon_r$&$\epsilon_a$ &  $\Delta $ \\
\hline
Penetrable Spheres & PS  &$>0$ &  $0$&$>0$       \\
Penetrable Square-Well & PSW  & $>0$ &  $>0$&$>0$       \\
Sticky Penetrable Spheres &  SPS  & $>0$ & $\to +\infty$&$\to 0$       \\
Penetrable Square-Shoulder &  PSS  &$>0$ & $<0$ &$>0$       \\
Hollow Penetrable Spheres &  HPS  & $0$ & $<0$&$>0$       \\
Hollow Hard Spheres & HHS  &$0$ &$\to - \infty$&$>0$       \\
\end{tabular}
\end{ruledtabular}
\end{table}

In all previous cases, we have tacitly assumed $\epsilon_a>0$. In
principle, however, nothing prevents to consider the opposite
 {case $\epsilon_a <0$ (what implies $\gamma_a<0$)}. In this
case the PSW potential gives rise to an interesting class of models,
at least from an academic point of view, with two  {positive}
energy scales ($\epsilon_r$ and $|\epsilon_a|$). If
$\epsilon_r>|\epsilon_a|$, we get a purely repulsive potential that
could be called the penetrable square-shoulder (PSS) model
 {(see Fig.\ \ref{potential})}. A peculiar situation occurs if
$\epsilon_r<|\epsilon_a|$: when two particles approach they
have first to overcome the barrier $|\epsilon_a|$ at
$r=\sigma+\Delta$; once done, they feel an attractive well of depth
$|\epsilon_a|-\epsilon_r$ for $r<\sigma$. Thus the potential is
attractive for short distances and repulsive for larger distances.
The simplest version of models with $\epsilon_a<0$ and
$\epsilon_r<|\epsilon_a|$ corresponds to $\epsilon_r=0$,    which
will be referred to as the hollow penetrable-sphere (HPS) model (see
Fig.\ \ref{potential}). If, in addition, the limit
$|\epsilon_a|\to\infty$ is taken, one gets  an \emph{athermal}
potential that will be referred to as the hollow hard-sphere (HHS)
model since the particles look like hard spheres of diameter
$\sigma+\Delta$ with a ``hole'' of diameter $\sigma$ inside. If two
particles are separated by a distance larger than $\sigma+\Delta$,
they behave as hard spheres and never know about their holes. On the
other hand, if the separation between them is smaller than $\sigma$,
they can never separate a  distance larger than $\sigma$. In the HHS
model, $\gamma_a \to -1$ and $\gamma_r \to 0$, so that the functions
$y_2(r)$ and $g_2(r)$ are well defined  {(see below)}. In Section
\ref{subsec:PY_HNC} we will discuss the results for representative
values of the parameters.  A summary of the penetrable models
treated in this paper, along with the  corresponding values for the
values for $\epsilon_r$, $\epsilon_a$, and $\Delta$ characterizing
them, is reported in Table \ref{tab:models}.
\section{ {Basic formalism for exact properties of nearest-neighbor potentials}}
\label{sec:formalism}

\subsection{ {General scheme}}
The great advantage of dealing with one-dimensional models is that
they are usually amenable to exact solutions, at least in the limit
of sufficiently short range interactions \cite{Lieb66}. The
trade-off is of course the fact that these models do not have phase
transitions. In the context of fluids, this translates into the fact
that there exist exact solutions for the HS, SHS, and SW models
\cite{Salsburg53,Feynman72,Seaton86,Heying04,Kikuchi55,Elkoshi85,
Corti98}. The same formalism allows one to tackle
non-nearest-neighbor one-dimensional fluids \cite{Malijevsky06} thus
leading to an approximate solution. Let us recall the main results
of this approach, referring to Ref.\ \cite{Heying04} for a
self-contained treatment. The main quantity to be computed is the
Laplace transform of the Boltzmann factor $e^{-\beta \phi(r)}$:
\begin{eqnarray}
\widetilde{\Omega}\left(s\right) &=& \int_{0}^{\infty} ds ~e^{-s r}
e^{-\beta \phi\left(r\right)}.
\label{formalism:eq1}
\end{eqnarray}
This is directly related to the Laplace transform of the radial
distribution function $g(r)$,
\begin{eqnarray}
\widetilde{G}\left(s\right)=\int_{0}^{\infty} ds~e^{-s r}
g\left(r\right),
\label{formalism:eq2}
\end{eqnarray}
where $\rho=N/L$ is the density of the one-dimensional fluid.
 {The relation is (see Refs.\ \cite{Santos07a,Heying04} for
details)}
\begin{equation}
\widetilde{G}\left(s\right)= \frac{1}{\rho} \left[ \frac
{\widetilde{\Omega}\left(\xi\right)}{\widetilde{\Omega}\left(s+\xi\right)}-1\right]^{-1}.
\label{formalism:eq3}
\end{equation}
 {Here  $\xi$ is a solution of the equation}
\begin{equation}
\left. \frac{\partial \ln \widetilde{\Omega}\left(s\right)}{\partial
s} \right|_{s=\xi} =- \frac{1}{\rho}.
\label{formalism:eq4}
\end{equation}
Finally, the equation of state (EOS) (and hence the whole
thermodynamics) can be cast into the very simple form
\begin{eqnarray}
\beta P &=& {\xi},
\label{formalism:eq5}
\end{eqnarray}
 {where $P$ is the pressure.}

In practice, the scheme goes as follows. Evaluate
$\widetilde{\Omega}(s)$ from the Boltzmann factor by a Laplace
transform, Eq.\ (\ref{formalism:eq1}); solve for $\xi$ from Eq.\
(\ref{formalism:eq4}); insert the result into Eq.\
(\ref{formalism:eq3}); invert the Laplace transform
(\ref{formalism:eq2}) to obtain $g(r)$ and, in parallel, compute the
EOS from (\ref{formalism:eq5}).

As a final remark, we anticipate that, when dealing with
discontinuous potentials (or Boltzmann factors), it is convenient to
introduce the cavity function $y(r)$ which is related to the radial
distribution function $g(r)$ and the pair potential $\phi(r)$ by the
general relation
\begin{eqnarray}
g\left(r\right) &=& e^{-\beta \phi\left(r\right)} y\left(r\right)
\label{formalism:eq6}.
\end{eqnarray}
Moreover it can be expanded in powers of the density
\begin{eqnarray}
y\left(r\right) &=& 1+\sum_{n=1}^\infty\rho^n y_n\left(r\right).
\label{formalism:eq7}
\end{eqnarray}
In principle, the knowledge of all $y_n$ coefficients provides the
exact solution to the cavity function $y(r)$ (provided that the
above series converges) and hence to the problem. This also allows
us to assess the reliability of well known approximations involving
the direct correlation function $c(r)$ and the  {cavity
function}  \cite{Hansen86}, such as the Percus-Yevick (PY) closure
\begin{equation}
c(r)=f(r)y(r)
\label{formalism:eq8}
\end{equation}
and the hypernetted chain (HNC) closure
\begin{equation}
c(r)=f(r)y(r)+y(r)-1 -\ln y(r).
\label{formalism:eq9}
\end{equation}

\subsection{ {Exact solution of the SHS model in
one dimension}}
\label{sec:exactshs}
Let us particularize the above procedure to derive the exact
solution of Baxter's SHS model in one dimension. Starting from the
Boltzmann factor
\begin{eqnarray}
e^{-\beta \phi\left(r\right)} &=& \Theta\left(r-\sigma\right) +
\alpha \sigma \delta_{+}\left(r-\sigma\right),
\label{exactshs:eq1}
\end{eqnarray}
its Laplace transform (\ref{formalism:eq1}) yields
\begin{eqnarray}
\widetilde{\Omega}\left(s\right) &=& \left(\alpha \sigma +
\frac{1}{s} \right) e^{-s \sigma}\label{exactshs:eq2}
\end{eqnarray}
Equation (\ref{formalism:eq4}) can then be arranged to  {get the
following quadratic equation}
\begin{eqnarray}
\xi^2 \sigma^2\alpha \left(1-\rho \sigma\right)+
\xi\sigma\left(1-\rho\sigma \right)-\rho \sigma=0.
\label{exactshs:eq3}
\end{eqnarray}
 {Its physical solution is}
\begin{eqnarray}
\xi =
\frac{\sqrt{1+4\alpha\rho\sigma/(1-\rho\sigma)}-1}{2\alpha\sigma},
\label{exactshs:eq4}
\end{eqnarray}
 {which can be substituted into Eq.\ (\ref{formalism:eq5}) to
give}
\begin{eqnarray}
\frac{\beta P}{\rho}
=\frac{\sqrt{1+4\alpha\rho\sigma/(1-\rho\sigma)}-1}{2\alpha\rho\sigma},
\label{exactshs:eq5}
\end{eqnarray}
which represents the EOS for this system. In order to get the
exact radial distribution function, we exploit Eq.\
(\ref{formalism:eq3}) to get
\begin{eqnarray}
\widetilde{G}\left(s\right)&=&\frac{1}{\rho}\sum_{n=1}^\infty
\left[\frac{\widetilde{\Omega}(s+\xi)}{\widetilde{\Omega}(\xi)}\right]^n\nonumber\\
&=&\frac{1}{\rho}\sum_{n=1}^{\infty} \frac{\left(\alpha \sigma +
\frac{1}{s+\xi}\right)^n e^{-n s \sigma}}{\left(\alpha\sigma +
\frac{1}{\xi}\right)^n }.
\label{exactshs:eq6}
\end{eqnarray}
We can now use the binomial theorem to expand $(\alpha \sigma +
\frac{1}{s+\xi})^n$ and invert the Laplace transform
(\ref{formalism:eq2}) term by term by using the residue theorem, to
obtain
\begin{eqnarray}
g\left(r\right)&=&\sum_{n=1}^{\infty}\psi_n(r-n\sigma)\Theta(r-n\sigma),\\
\psi_n(r)&=&\frac{1}{\rho}\frac{1}{(\alpha\sigma+1/\xi)^n }
\left[(\alpha\sigma)^n\delta_+(r)+\sum_{k=1}^{n}\binom{n}{k}
(\alpha\sigma)^{n-k}\frac{r^{k-1}e^{-\xi r}}{(k-1)!} \right],
\label{exactshs:eq7}
\end{eqnarray}
which is the correct result found in Ref.\ \cite{Seaton86} with a
different method.
\section{Exact properties of the SPS model}
\label{sec:exactsps}
Next we turn our attention to the corresponding penetrable SPS
counterpart. Following Ref.\ \cite{Malijevsky06}, the basic idea
hinges on deducing the exact low-density orders of the SPS model
from those of the SHS, which can be evaluated exactly. Each term
$y_n(r)$ can be represented as a sum of suitable diagrams, whose
forms for $y_1(r)$ and $y_2(r)$ were given in Ref.\
\cite{Malijevsky06} and will not be repeated here.  {Each bond
in the diagrams corresponds to a Mayer function $f(r)$ and the ones
for the SPS and SHS models are related by Eq.\ (\ref{sps:eq3})}. For
the SHS model previously discussed,  {the exact cavity function
$y(r)$ does not have a Dirac delta function at  $r=\sigma$ and the
regular part is continuous at that point}.  {None of these two
properties is any longer true for the SPS model}, as further
elaborated below. Here and in the following we set $\sigma=1$ for
simplicity.

 {The result is}
\begin{eqnarray}
\label{exactsps:eq3a}
y_1^{(\SPS)}\left(r\right) &=& \gamma_r^2 y_{1}^{(\SHS)}\left(r\right), \\
y_2^{(\SPS)}\left(r\right) &=& \gamma_r^3 y_{2A}^{(\SHS)}
\left(r\right)+ 2\gamma_r^4 y_{2B}^{(\SHS)}\left(r\right) +
\frac{\gamma_r^4}{2} y_{2C}^{(\SHS)}\left(r\right)+
\frac{\gamma_r^5}{2} y_{2D}^{(\SHS)}\left(r\right),
\label{exactsps:eq3b}
\end{eqnarray}
where the first-order density term is
\begin{eqnarray}
y_1^{(\SHS)}\left(r\right)&=&\left(2-r-2\alpha\right)\Theta\left(2-r\right)+
\alpha^2
\left[2\delta_+\left(r\right)+\delta_+\left(r-2\right)\right].
\label{exactsps:eq4}
\end{eqnarray}
Note that this has a delta singularity at $r=0$ and is continuous at
$r=1$. For the second order in density we have
\begin{eqnarray}
y_{2A}^{(\SHS)}\left(r\right)&=&\left[-\left(3-r^2\right)+
6\alpha\left(1-\alpha\right)\right] \Theta\left(1-r\right)
\nonumber\\
&&+\left[-\frac{1}{2}\left(3-r\right)^2+3\alpha\left(3-\alpha-r\right)\right]
\left[\Theta\left(3-r\right)-\Theta\left(1-r\right)\right]
\nonumber\\
&&+\alpha^3\left[3\delta_+\left(r-1\right)+\delta_+\left(r-3\right)\right]~,
\label{y2ASHS}\\
y_{2B}^{(\SHS)}\left(r\right)&=&\left[\frac{1}{2}\left(6-2r-r^2\right)-
\alpha\left(6-6\alpha-r\right)\right] \Theta\left(1-r\right)\nonumber\\
&&+\left[\frac{1}{2}\left(2-r\right)\left(4-r\right)-\alpha\left(8-4\alpha-3r
\right)\right]
\left[\Theta\left(2-r\right)-\Theta\left(1-r\right)\right]
\nonumber\\
&&+\alpha^2\left(1-2\alpha\right)\left[2\delta_+\left(r\right)+
\delta_+\left(r-2\right)\right]
-2\alpha^3\delta_+\left(r-1\right)~,
\label{y2BSHS}\\
y_{2C}^{(\SHS)}(r)&=&[y_1^{(\SHS)}(r)]^2~,
\label{y2CSHS}\\
y_{2D}^{(\SHS)}(r)&=&[-(3-2r)+2\alpha(3-3\alpha-r)]\Theta(1-r)\nonumber\\
&&+[-(2-r)^2+4\alpha(2-\alpha-r)][\Theta(2-r)-\Theta(1-r)]\nonumber\\
&&+2\alpha^3[6\delta_+(r)+\delta_+(r-1)+3\delta_+(r-2)]-
\alpha^4[4\delta_+^2(r)+\delta_+^2(r-2)]~.
\label{exactsps:eq5}
\end{eqnarray}
 {The functions
\eqref{y2ASHS}--\eqref{exactsps:eq5} present some peculiar
properties. In particular, (i) the regular parts of
$y_{2A}^{(\SHS)}(r)$, $y_{2B}^{(\SHS)}(r)$, and $y_{2D}^{(\SHS)}(r)$
are discontinuous at $r=1$; (ii) $y_{2A}^{(\SHS)}(r)$,
$y_{2B}^{(\SHS)}(r)$, and $y_{2D}^{(\SHS)}(r)$ have a delta
singularity at $r=1$; and (iii) $y_{2C}^{(\SHS)}(r)$ and
$y_{2D}^{(\SHS)}(r)$ present delta-square singularities at $r=0$ and
$r=2$. However, these three classes of singularities cancel out when
setting $\gamma_r=1$ in Eq.\ \eqref{exactsps:eq3b} to obtain the
total second-order function $y_2^{(\SHS)}(r)$ \cite{note2}. On the
other hand, since for SPS $y_{2A}^{(\SHS)}(r)$,
$y_{2B}^{(\SHS)}(r)$, $y_{2C}^{(\SHS)}(r)$, and $y_{2D}^{(\SHS)}(r)$
are weighed by different powers of $\gamma_r$ (3, 4, 4, and 5,
respectively), the corresponding exact second-order cavity function
 is discontinuous at $r=1$ and has a delta singularity
at $r=1$ and  delta-square singularities at $r=0$ and $r=2$. The
delta singularity at $r=1$ is responsible for a diverging fourth
virial coefficient of the SPS model (see below).}

\section{Exact properties of the PSW model}
\label{sec:exactpsw}
\subsection{Calculation of $y_1$ and $y_2$}
\label{subsec:y1_y2}
As already mentioned, the SPS model suffers from the same drawbacks
as the original SHS model  {plus some additional ones, so that it}
can hardly be regarded as a sound model in higher dimensions.
However it has served as a test bench for  analytical techniques.
Armed by these tools we can now tackle the more difficult PSW model
which has the SW fluid as a reference model. We recall that the
latter does not have an exact solution in higher dimensions but it
is amenable to an exact treatment in one dimension \cite{Heying04}.
The discussion follows closely the route already introduced for the
SPS model, namely the density expansion, Eq.\ (\ref{formalism:eq7}).
The radial distribution function $g(r)$ is related to the cavity
function $y(r)$ by Eq.(\ref{formalism:eq6}) which with the help of
Eqs. (\ref{psw:eq2}) and (\ref{psw:eq3}) yields
\begin{eqnarray}
g\left(r\right)&=&\left\{
\begin{array}{ll}
\left(1-\gamma_r\right)y\left(r\right), & r<1, \\
\left(1+\gamma \gamma_r\right)y\left(r\right), & 1<r<1+\Delta ,\\
y\left(r\right)   ,   & r>1+\Delta.
\end{array}\right. \label{exactpsw:eq1}
\end{eqnarray}
As in the SPS model, the cavity function can be exactly computed up
to second order in density, this time by reducing the problem to the
solution of the SW model.

The first order term reads ($\Delta <1$)
\begin{eqnarray} \label{exactpsw:eq2}
y_1\left(r\right)=\gamma_r^2 \begin{cases}
2\left(1+\gamma^2\Delta\right)-
r\left(1+2\gamma+2\gamma^2\right),&0\leq r\leq\Delta,\\
2-2 \gamma\Delta-r,  & \Delta \leq r\leq 2, \\
 \gamma\left(2+\gamma\right)\left(r-2\right)-2\gamma\Delta ,
& 2\leq r\leq 2+\Delta,  \\
 \left(2+2\Delta -r\right)\gamma^2, & 2+\Delta \leq r\leq 2+2\Delta,  \\
 0, & 2+2\Delta \leq r.
\end{cases}
\end{eqnarray}
The second order can be reduced to the calculation of the
corresponding diagrams of the SW model as anticipated. We find
\begin{eqnarray}
y_2\left(r\right) &=& \gamma_r^3 y_{2A}^{(\SW)} \left(r\right) + 2
\gamma_r^4 y_{2B}^{(\SW)} \left(r\right) + \frac{\gamma_r^4}{2}
y_{2C}^{(\SW)}\left(r\right) + \frac{\gamma_r^5}{2} y_{2D}^{(\SW)},
\left(r\right)
\label{exactpsw:eq3}
\end{eqnarray}
where the explicit calculation of the various terms is described in
Appendix \ref{app:sw}  {and is given by Eqs.\ \eqref{appsw:eq2},
\eqref{appsw:eq3}, \eqref{appsw:eq5}, and \eqref{appsw:eq9}. It can
be checked that these expressions reduce to those of the SPS model,
Eqs.\ \eqref{y2ASHS}--\eqref{exactsps:eq5}, in the limit
$\gamma\to\infty$ and $\Delta\to 0$ with
$\alpha=\gamma\Delta/\sigma=\text{const}$.}
\subsection{Computation of   {$B_2$, $B_3$, and $B_4$}}
\label{subsec:virial}
The EOS can be obtained from the knowledge of the radial
distribution function $g(r)$ through a number of routes. The most
common ones are the virial route
\begin{eqnarray}
\frac{\beta P}{\rho} &\equiv&Z\left(\rho,\beta\right) = 1 + 2^{d-1}
v_d \rho \int_{0}^{\infty} dr~ r^d y\left(r\right)
\frac{\partial}{\partial r} f\left(r\right),
\label{virial:eq1}
\end{eqnarray}
the compressibility route
\begin{eqnarray}
\left(\beta \frac{\partial P}{\partial \rho} \right)^{-1} \equiv
\chi\left(\rho,\beta\right) = 1+2^d d v_d \rho \int_{0}^{\infty}
dr~r^{d-1} \left[g\left(r\right)-1\right],
\label{virial:eq2}
\end{eqnarray}
and the energy route
\begin{eqnarray}
\frac{U}{N} &\equiv& u\left(\rho,\beta\right)= \frac{d}{2 \beta}
\left[ 1+ 2^d v_d \rho \beta \int_{0}^{\infty} dr~r^{d-1}
\phi\left(r\right) g\left(r\right) \right],
\label{virial:eq3}
\end{eqnarray}
where $d$ is the dimensionality of the system and $v_d=(\pi/4)^{d/2}/\Gamma(1+d/2)$
is the volume of a $d$-dimensional sphere of unit diameter. Thermodynamic consistency
for the \textit{exact} $g(r)$ requires the three routes to be completely equivalent
and hence
\begin{eqnarray} \label{virial:eq4}
\chi^{-1}\left(\rho,\beta\right) &=& \frac{\partial}{\partial \rho} \left[
\rho Z\left(\rho,\beta\right) \right], \\
\rho \frac{\partial}{\partial \rho} u\left(\rho,\beta\right) &=&
\frac{\partial}{\partial \beta} Z\left(\rho,\beta\right).
\label{virial:eq5}
\end{eqnarray}
For an \textit{approximate} $g(r)$, on the other hand, the consistency is
no longer granted and different routes (or combinations of them) may lead
to different results.

Let us specialize to the one-dimensional case of the PSW model,
where we have just derived the exact $g(r)$ up to second order in a
density expansion. Equations (\ref{virial:eq1})--(\ref{virial:eq3})
become, using the potential (\ref{psw:eq1}),
\begin{eqnarray}\label{virial:eq6}
Z\left(\rho,\beta\right) &=& 1+\rho \gamma_r \left[\left(1+\gamma\right) y\left(1\right)
-\gamma \left(1+\Delta\right) y\left(1+\Delta\right)\right], \\
\label{virial:eq7}
\chi\left(\rho,\beta\right) &=& 1+2 \rho \left\{ \int_{0}^{1} dr~\left[\left(1-\gamma_r\right)
y\left(r\right) -1 \right]+\int_{1}^{1+\Delta} dr~\left[\left(1+\gamma_r \gamma\right)
y\left(r\right) -1\right]+\int_{1+\Delta}^{+\infty} dr~\left[y\left(r\right)-1\right] \right\}, \\
\label{virial:eq8}
u\left(\rho,\beta\right)&=& \frac{1}{2\beta} + \rho \left[\epsilon_r
\left(1-\gamma_r\right) \int_{0}^{1} dr~y\left(r\right)- \epsilon_a
\left(1+\gamma_r \gamma\right) \int_{1}^{1+\Delta}
dr~y\left(r\right) \right].
\end{eqnarray}
Inserting the expansion (\ref{formalism:eq7}) for the cavity function $y(r)$
we find
\begin{eqnarray}
\label{virial:eq9}
Z&=& 1+ B_2 \rho +B_3 \rho^2 + B_4 \rho^3+ \cdots, \\
\label{virial:eq10}
\chi &=& 1+ \chi_2 \rho + \chi_3 \rho^2 + \chi_4 \rho^3 + \cdots, \\
\label{virial:eq11}
u &=& \frac{1}{2 \beta} + u_2 \rho + u_3 \rho^2 + u_4 \rho^3 +
\cdots.
\end{eqnarray}
Clearly Eq.\ (\ref{virial:eq9}) is the virial expansion for the
compressibility factor $Z$ whereas (\ref{virial:eq10}) and
(\ref{virial:eq11}) are the analogous expansions for the isothermal
compressibility $\chi$ and the energy per particle $u$. If the exact
coefficients $y_n$ appearing in Eq.\ (\ref{formalism:eq7}) are
known, the above three quantities provide the identical exact EOS.

On starting from the second-order values $B_2$, $\chi_2$, and $u_2$
one can obtain perturbatively higher orders term by term from the
knowledge of $y_n(r)$.  {The result can be cast into the form}
\begin{eqnarray}
\label{virial:eq12}
B_2 &=& \gamma_r \left(1-\gamma \Delta\right) ,\quad
\chi_2=-2B_2,\quad
u_2=\epsilon_r\left(1-\gamma_r\right)-\epsilon_a\left(1+\gamma_r
\gamma\right)\Delta,
\\
\label{virial:eq13}
B_n &=& \gamma_r \left[\left(1+\gamma\right) y_{n-2}\left(1\right)
-\gamma \left(1+\Delta\right) y_{n-2}\left(1+\Delta\right) \right],
\qquad n \ge 3,\\
\label{virial:eq15}
\chi_n&=&2\left[\left(1-\gamma\right)\int_{0}^{1} d r\, y_{n-2}\left(r\right)+
\left(1+\gamma_r \gamma \right)\int_{1}^{1+\Delta}
d r\,y_{n-2}\left(r\right)+\int_{1+\Delta}^\infty d r\, y_{n-2}\left(r\right)\right],
\qquad n\geq 3, \\
 \label{virial:eq17}
u_n&=&\epsilon_r\left(1-\gamma_r\right)\int_{0}^{1} d r\,
y_{n-2}\left(r\right)-\epsilon_a\left(1+\gamma_r \gamma\right)
\int_1^{1+\Delta} d r\, y_{n-2}\left(r\right),\qquad
n\geq 3.
\end{eqnarray}
 {Note that $B_n$ depends upon $y_{n-2}$ so that knowledge of
the exact $y_1$ and $y_2$ allows the computation of the exact virial
coefficients up to $B_4$. The third- and fourth-order results can be
obtained from Eqs.\ (\ref{exactpsw:eq2}) and (\ref{exactpsw:eq3}).
After some algebra, one gets}
\begin{eqnarray}
B_3&=&\gamma_r^3\left[1-\gamma \Delta\left(2-\Delta-2 \gamma
\Delta\right)\right],\label{virial:eq19}
\\ \label{virial:eq20}
\chi_3&=&4B_2^2-3B_3,
\\
u_3&=&\frac{\epsilon_r}{2}\gamma_r^2\left(1-\gamma_r\right)
\left[3-2\gamma \Delta\left(2-\Delta-\gamma \Delta\right)\right]-
\frac{\epsilon_a}{2}\gamma_r^2\left(1+\gamma_r \gamma
\right)\Delta\left(2-\Delta- 4 \gamma \Delta\right)\nonumber \\
&=&\frac{1}{2}\frac{\partial}{\partial\beta} B_3,\label{virial:eq21}
\end{eqnarray}
\begin{eqnarray}
B_4&=&-\frac{\gamma_r^6}{2}\left[1-\gamma\Delta\left(3-3\Delta-6\gamma\Delta+
\Delta^2+4\gamma
\Delta^2+3\gamma^2\Delta^2-\gamma^3\Delta^2\right)\right]\nonumber
\\ \nonumber
&&+\frac{\gamma_r^5}{2}\left[7-\gamma \Delta \left(21-15\Delta-36
\gamma \Delta+ 3\Delta^2+16\gamma
\Delta^2+16\gamma^2\Delta^2-4\gamma^3\Delta^2 \right)\right]
\\ \label{app_exact:eq1}&&
-\frac{\gamma_r^4}{2} \left[4-\gamma\Delta\left(12-6\Delta-18\gamma
\Delta+\Delta^2+3\gamma \Delta^2+
3\gamma^2\Delta^2-3\gamma^3\Delta^2\right)\right],\\
\chi_4&=&-4\left(2B_2^3-3B_2B_3+B_4\right),\\
\label{app_exact:eq2}
u_4&=&\frac{1}{3}\frac{\partial}{\partial\beta}B_4.
\label{app_exact:eq3}
\end{eqnarray}
 {The three routes provide consistently identical results for
$B_3$ and $B_4$, i.e., the relations (\ref{virial:eq4}) and
(\ref{virial:eq5}) are verified, as they should. In the energy case
the following identity is needed:}
\begin{eqnarray}
\frac{\partial}{\partial\beta}&=&\frac{\partial
\gamma_r}{\partial\beta}\frac{\partial}{\partial
\gamma_r}+\frac{\partial
\gamma}{\partial\beta}\frac{\partial}{\partial \gamma}\nonumber \\
&=&\epsilon_r\left(1-\gamma_r\right)\frac{\partial}{\partial
\gamma_r}+\frac{\epsilon_a \left(1+\gamma_r \gamma\right)-\epsilon_r
\gamma\left(1-\gamma_r\right)}{\gamma_r}\frac{\partial}{\partial
\gamma}.
\label{virial:eq18}
\end{eqnarray}

 {Equation \eqref{app_exact:eq1} gives the exact fourth virial
coefficient as a function of the three relevant parameters of the
PSW model, namely $\gamma_r$, $\gamma$, and $\Delta$. The results
for the PS and SW models are recovered as}
\begin{eqnarray}
\lim_{\epsilon_a\to 0}B_4=\lim_{\epsilon_a\to
-\epsilon_r}\frac{B_4}{(1+\Delta)^3}=\gamma_r^4\left(-\frac{\gamma_r^2}{2}+
\frac{7\gamma_r}{2}-2\right),
\label{app_exact:eq4}
\end{eqnarray}
\begin{eqnarray}
\lim_{\epsilon_r\to \infty}B_4=1-\gamma
\Delta\left(3-3\Delta-6\gamma \Delta+\frac{1}{2}\Delta^2+
\frac{9}{2}\gamma \Delta^2 +5\gamma^2\Delta^2\right),
\label{app_exact:eq5}
\end{eqnarray}
 {respectively. On the other hand, while  $B_2$ and $B_3$ are
well defined in the SPS limit ($\gamma \to\infty$ and $\Delta\to 0$
with $\alpha=\gamma \Delta=\text{finite}$)  [see Eqs.\
(\ref{virial:eq12}) and (\ref{virial:eq19})], the presence of the
terms $\gamma^3\Delta^2$ in Eq.\ (\ref{app_exact:eq1}) implies that
$B_4\to\infty$ in the SPS model. Equation \eqref{virial:eq13} shows
that this is a direct consequence of the divergence of
$y_2^{(\SPS)}(r)$ at $r=1$. However,
$y_2^{(\SHS)}(1)=\text{finite}$, so that $B_4$ is well defined in
the SHS model ($\gamma_r=1$), as shown by Eq.\
(\ref{app_exact:eq5}).}

 The second, third, and fourth virial coefficients for the PSS
model ($\epsilon_a<0$) are still given by Eqs.\ (\ref{virial:eq12}),
(\ref{virial:eq19}), and \eqref{app_exact:eq1}, except that
$\gamma<0$. In the  case of the HPS model ($\epsilon_a<0$ and
$\epsilon_r\to 0$ or, equivalently, $\gamma_a<0$,
$\gamma=\gamma_a/\gamma_r$, and $\gamma_r\to 0$), one gets
\begin{eqnarray}
\lim_{\epsilon_r\to 0}B_2=-\gamma_a\Delta,\quad\lim_{\epsilon_r\to
0}B_3=0,\quad \lim_{\epsilon_r\to
0}B_4=-\frac{3}{2}\gamma_a^4\Delta^3.
\label{app_exact:eq6}
\end{eqnarray}
The special case of the HHS model is obtained by further
taking the limit $\epsilon_a\to\-\infty$ ($\gamma_a\to -1$).

\section{ {Some illustrative cases and comparison with the PY and HNC
approximations}}
\label{subsec:PY_HNC}
\begin{figure}
\includegraphics[width=.4\columnwidth]{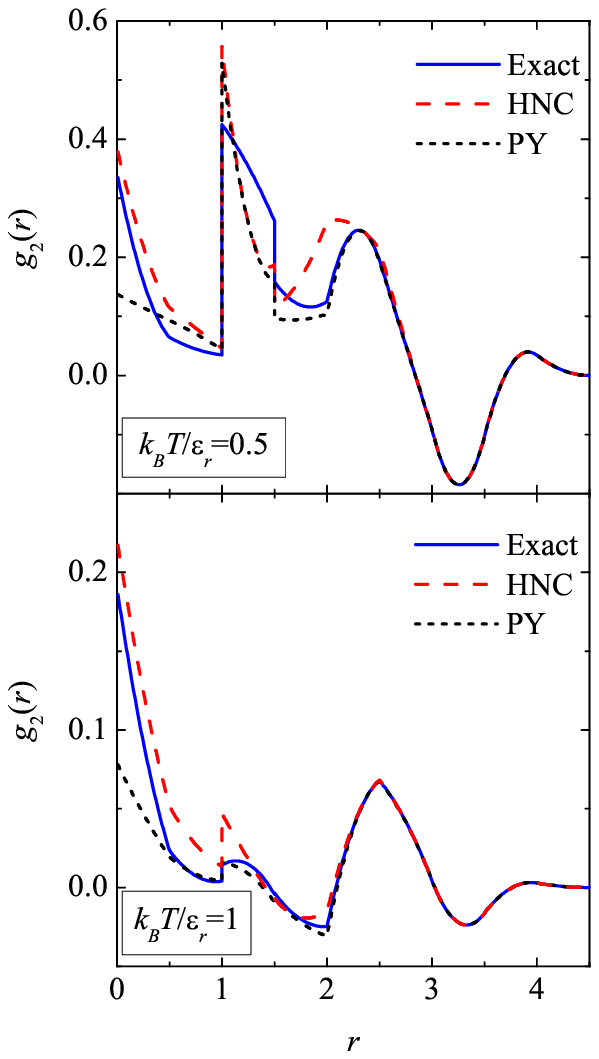}
\caption{ {(Color online) Plot of the second-order radial
distribution function, $g_2(r)$, for a PSW model with
$\epsilon_a/\epsilon_r=0.25$, $\Delta=0.5$, and
$k_BT/\epsilon_r=0.5$ (top panel) and $k_BT/\epsilon_r=1$ (bottom
panel). The solid, long-dashed, and short-dashed lines correspond to
the exact result, the HNC approximation, and the PY approximation,
respectively}.\label{g2_PSW}}
 \end{figure}
\begin{figure}
\includegraphics[width=.4\columnwidth]{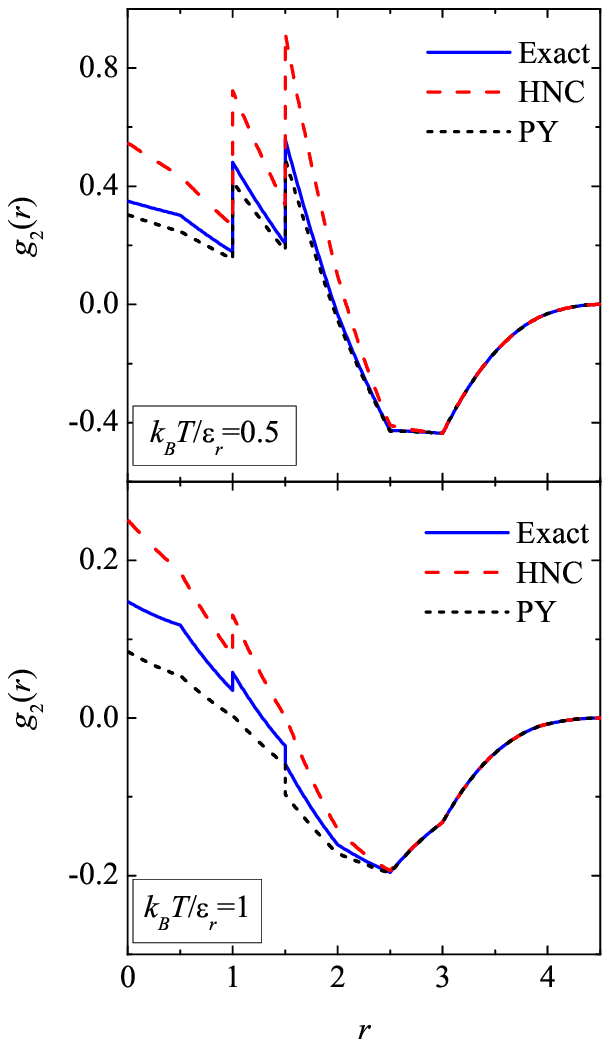}
\caption{ {(Color online) Plot of the second-order radial
distribution function, $g_2(r)$, for a PSS model with
$\epsilon_a/\epsilon_r=-0.5$, $\Delta=0.5$, and
$k_BT/\epsilon_r=0.5$ (top panel) and $k_BT/\epsilon_r=1$ (bottom
panel). The solid, long-dashed, and short-dashed lines correspond to
the exact result, the HNC approximation, and the PY approximation,
respectively}.\label{g2_PSS}}
 \end{figure}
\begin{figure}
\includegraphics[width=.4\columnwidth]{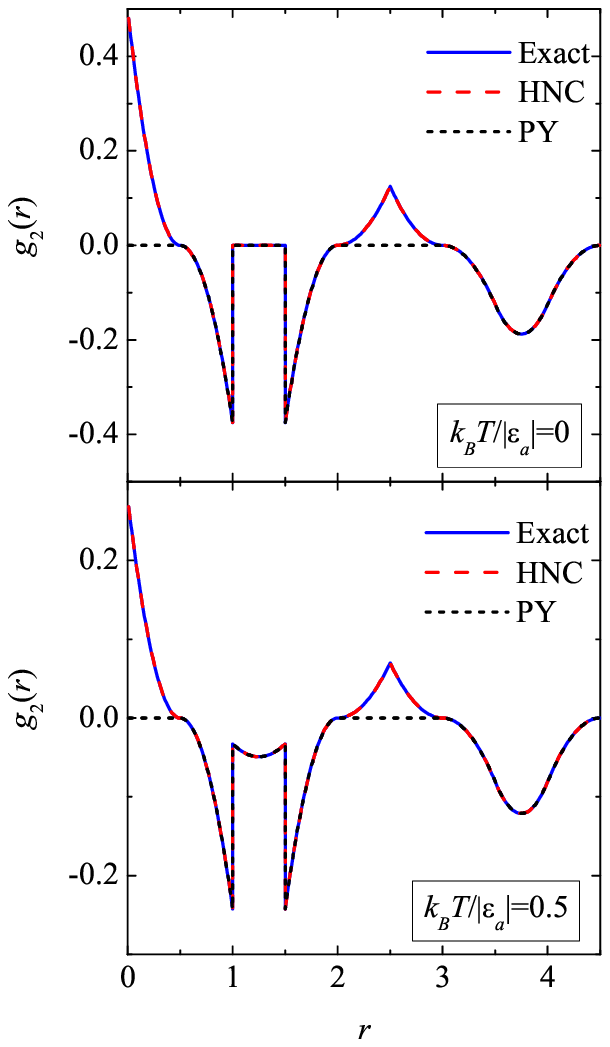}
\caption{ {(Color online) Plot of the second-order radial
distribution function, $g_2(r)$, for the HPS model with
 $\Delta=0.5$ and  $k_B T/|\epsilon_a|=0.5$ (bottom panel)
and $k_B T/|\epsilon_a|=0$ (top panel, corresponding to the HHS
model). The solid, long-dashed, and short-dashed lines correspond to
the exact result, the HNC approximation, and the PY approximation,
respectively. Note that the HNC approximation provides the exact
result in the HPS model.}\label{g2_HPS}}
 \end{figure}
  The approximate character of  a given closure can be typically inferred by looking
at $g_2(r)$ along with the corresponding  fourth virial coefficient
$B_4$. Being coefficients in a density expansion, both can be either
positive or negative. We now plot the exact $g_2(r)$ and $B_4$
for some illustrative cases and  compare them with the PY and HNC
theories (\ref{formalism:eq8}) and (\ref{formalism:eq9}). The PY and
HNC results corresponding to Eq.\ (\ref{exactpsw:eq3}) are
\cite{Hansen86}
\begin{eqnarray}
y_{2}^{\text{PY}}\left(r\right)=\gamma_r^3y_{2A}^{(\text{SW})}\left(r\right)+
2\gamma_r^4y_{2B}^{(\text{SW})}\left(r\right),
\label{PY_HNC:eq2}
\end{eqnarray}
\begin{eqnarray}
y_{2}^{\text{HNC}}\left(r\right)&=&\gamma_r^3y_{2A}^{(\text{SW})}\left(r\right)
+2\gamma_r^4y_{2B}^{(\text{SW})}\left(r\right)+\frac{\gamma_r^4}{2}y_{2C}^{(\text{SW})}\left(r\right).
\label{PY_HNC:eq1}
\end{eqnarray}
 {Comparison with Eq.\ \eqref{exactpsw:eq3} shows that the HNC
theory neglects $y_{2D}^{(\text{SW})}\left(r\right)$ and the PY
theory neglects, in addition, $y_{2C}^{(\text{SW})}\left(r\right)$.
As a consequence, the expression for the fourth virial coefficient
in the PY and HNC approximations depends on the thermodynamic route.
The corresponding results can be found in Appendix \ref{app:B4}.}

 {Let us start with $g_2(r)$. As a prototypical PSW system we
have chosen  $\epsilon_a/\epsilon_r=0.25$ and $\Delta=0.5$. Figure
\ref{g2_PSW} shows $g_2(r)$ for $k_BT/\epsilon_r=0.5$ and
$k_BT/\epsilon_r=1$}. It can be observed that the HNC and PY tend to
overestimate and underestimate, respectively, the values of $g_2(r)$
in the overlapping region $r<1$. This is due to the fact that
$y_{2D}^{(\SW)}(r)$ is generally negative in the region $r<1$, while
$y_{2C}^{(\SW)}(r)$ is positive definite and larger than the
magnitude of $y_{2D}^{(\SW)}(r)$.  Inside the well ($1<r<1+\Delta$)
the PY and HNC curves practically coincide at $k_BT/\epsilon_r=0.5$,
both being rather inaccurate, while at the higher temperature
$k_BT/\epsilon_r=1$ the PY prediction is quite good.
Moreover, the PY theory is a better approximation than the HNC
theory for $r>2$. This is explained by the fact that
$y_{2C}^{(\SW)}(r)+y_{2D}^{(\SW)}(r)=0$ in the region $r>2$, so that
$g_2^{\text{PY}}(r)$ coincides with the exact $g_2(r)$ for $r>2$ in
the case of the SW model ($\gamma_r=1$). If $\gamma_r<1$ the
combination $y_{2C}^{(\SW)}(r)+ \gamma_r y_{2D}^{(\SW)}(r)$ does not
vanish for $r>2$ but is still rather small for the cases of Fig.\
\ref{g2_PSW}. For $r>2+2\Delta$, both $g_2^{\text{HNC}}(r)$ and
$g_2^{\text{PY}}(r)$ become exact since $y_{2C}^{(\SW)}(r)$ and
$y_{2D}^{(\SW)}(r)$ vanish in that region.

Figure \ref{g2_PSS}  depicts the function  $g_2(r)$ for a
representative case of the PSS model (see Section
\ref{subsec:other}).  Most of the preceding comments in connection
with Fig.\ \ref{g2_PSW} apply here as well.  Finally, the function
$g_2(r)$ corresponding to the HPS model
 is shown in Fig.\
\ref{g2_HPS} for $\Delta=0.5$ and two temperatures:
$k_BT/|\epsilon_a|=0$ and $k_BT/|\epsilon_a|=0.5$. Note that the
zero-temperature case is equivalent to the HHS limit. It is
interesting to note that the  curves corresponding to both
temperatures are quite similar, except for a change of scale. In
the HPS model the HNC theory gives the exact $g_2(r)$
because, for large $|\gamma|$, $y_{2D}^{(\text{SW})}(r)$
scales with $\gamma^4$, while it has a weight $\gamma_r^5$ and so
does not contributes to $y_2(r)$. Similarly,
$y_{2B}^{(\text{SW})}(r)$ scales with $\gamma^3$ and so it does not
contribute to $y_2(r)$ either. On the other hand,
$\gamma_r^4y_{2C}^{(\text{SW})}(r)$ is different from zero in the
regions $0\leq r\leq \Delta$ and $2\leq r\leq 2+2\Delta$ and it is
there where the PY theory fails, yielding $g_2^{\text{PY}}(r)=0$.

\begin{figure}
\includegraphics[width=.4\columnwidth]{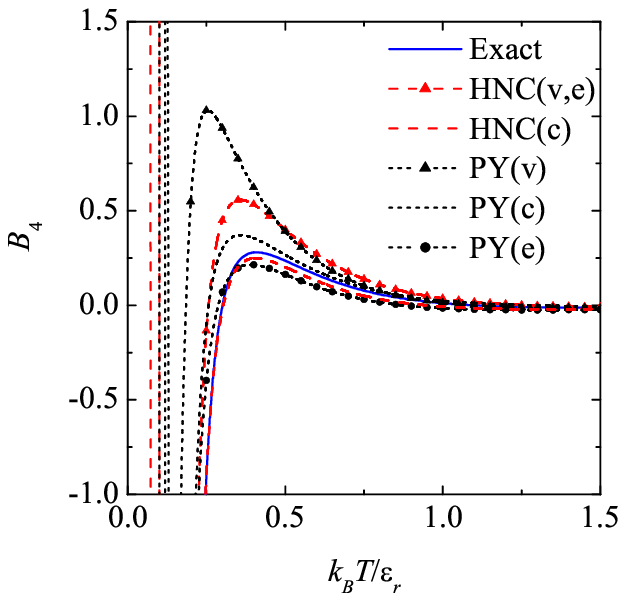}
 \caption{ {(Color online) Plot of the fourth virial coefficient, $B_4$, as a function of $k_BT/\epsilon_r$ for a PSW
 model with
$\epsilon_a/\epsilon_r=0.25$ and $\Delta=0.5$. The solid, dashed,
and dotted lines correspond to the exact result, the HNC
approximation (virial/energy and compressibility routes), and the PY
approximation (virial, compressibility, and energy routes),
respectively.}\label{B4_PSW}}
 \end{figure}
\begin{figure}
\includegraphics[width=.4\columnwidth]{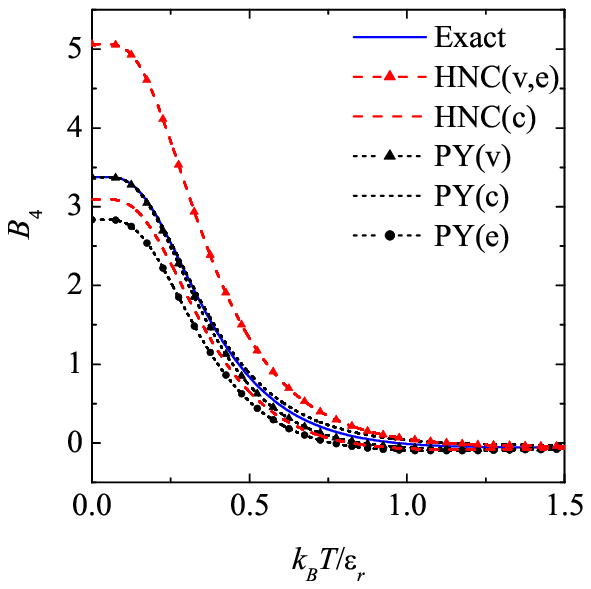}
 \caption{ {(Color online) Plot of the fourth virial coefficient, $B_4$, as a function of $k_BT/\epsilon_r$ for a
 PSS
 model with
$\epsilon_a/\epsilon_r=-0.5$ and $\Delta=0.5$. The solid, dashed,
and dotted lines correspond to the exact result, the HNC
approximation (virial/energy and compressibility routes), and the PY
approximation (virial, compressibility, and energy routes),
respectively.}\label{B4_PSS}}
 \end{figure}
\begin{figure}
\includegraphics[width=.4\columnwidth]{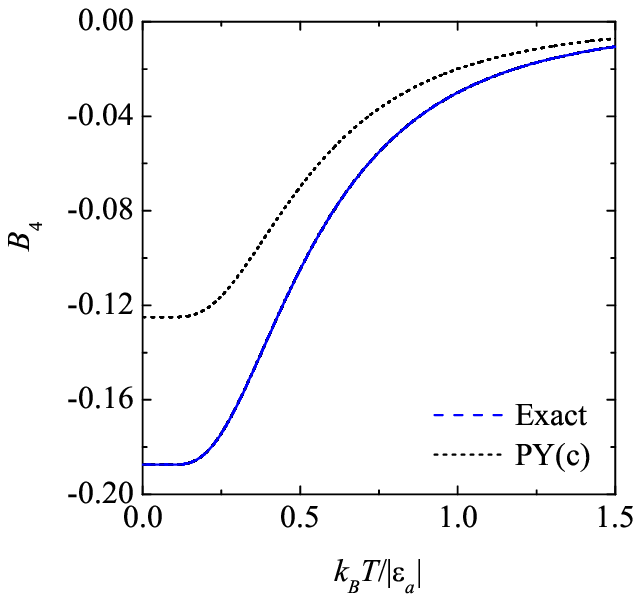}
 \caption{(Color online)
Plot of the fourth virial coefficient, $B_4$, as a function of
$k_BT/|\epsilon_a|$ for a HPS model with $\Delta=0.5$. The dashed
line corresponds to the exact result whereas the dotted line
correspond to the PY approximation (compressibility route).
\label{B4_HPS}}
 \end{figure}
In order to have a feeling of the behavior of the exact $B_4$, we
now plot them for some representative values of the parameters.
Figure \ref{B4_PSW} shows  {the exact [see Eq.\
\eqref{app_exact:eq1}] and the approximate (see Appendix
\ref{app:B4})} values of the fourth virial coefficient as  functions
of temperature for the same PSW model as considered in Fig.\
\ref{g2_PSW}, i.e., the one defined by $\epsilon_a/\epsilon_r=0.25$
and $\Delta=0.5$. While the exact $B_4$ goes to $-\infty$ as $T\to
0$, the HNC and PY theories artificially predict a divergence to
$+\infty$. We can observe that the best agreement with the exact
curve corresponds to $B_4^{\text{HNC},c}$ up to
 {$k_BT/\epsilon_r\simeq 0.5$} and to $B_4^{\text{PY},c}$
thereafter.  {The worst behaviors correspond to
$B_4^{\text{HNC},v}=B_4^{\text{HNC},e}$ and $B_4^{\text{PY},v}$.}

The results for the PSS model considered in Fig.\ \ref{g2_PSS},
namely $\epsilon_a/\epsilon_r=-0.5$ and $\Delta=0.5$, are displayed
in Fig.\ \ref{B4_PSS}. For low temperatures this model reduces to
the HS model of diameter $1+\Delta$. It is found that
$B_4^{\text{PY},v}$ and, especially, $B_4^{\text{PY},c}$ present an
excellent agreement with the exact $B_4$.  {On the other hand,
the poorest performances are presented by
$B_4^{\text{HNC},v}=B_4^{\text{HNC},e}$ and $B_4^{\text{PY},e}$.}

 We have also evaluated $B_4$ for the HPS model at various values
of $k_B T/|\epsilon_a|$, as depicted in Fig.\ \ref{B4_HPS}, and
compared with the PY approximation (compressibility route). As
said before, the HNC theory becomes exact for the HPS model.
Interestingly, in this case both the virial and the energy routes
from the PY approximation yield exact results, even though
$y_2^\text{PY}(r)$ is not exact.

It is worthwhile noting that $B_4$ is not a monotonic function
of temperature in the PSW model (see Fig.\ \ref{B4_PSW}): it is
negative for low temperatures, reaches a positive maximum value at
an intermediate temperature, and then  decays, reaching a very small
negative minimum value at a certain temperature, and finally going
to zero from below. Although hardly apparent in Fig.\ \ref{B4_PSS},
the behavior of $B_4$ is also non-monotonic in the PSS model: it is
generally positive and decays as the temperature increases, but
eventually reaches a very small negative minimum value and
thereafter tends to zero from below. In contrast, the fourth virial
coefficient of the HPS model (see Fig.\ \ref{B4_HPS}) is negative
definite and monotonically increases with increasing temperature.

\section{Conclusions and outlook}
\label{sec:conclusions}
In this paper, we have introduced the PSW model and outlined a
number of exact results for this model in one dimension.  {The
potential contains two energy scales (the core barrier $\epsilon_r$
and the well depth $\epsilon_a$) and two length scales (the core
diameter $\sigma$ and the well width $\Delta$).} This model is a
variation of the widely used square-well one with a finite energy
barrier replacing the hard core. As such, this is not a
nearest-neighbor system and there exists no general approach leading
to an exact solution even in the one-dimensional case. In spite of
this we have  {been able to obtain the exact first few}
coefficients in the density expansions of the relevant structural
and thermodynamical properties. Specifically, we have computed both
the cavity and radial distribution functions up to second order in
density and the virial expansion up to fourth order.  {As a
stringent test of the calculations,} we have explicitly checked that
different routes to thermodynamics (virial, compressibility, and
energy) are consistent one another up to this order.

This model includes a variety of other models as special cases. By
taking the limit of infinitely narrow and deep well
($\epsilon_a\to\infty$, $\Delta\to 0$) we obtain the SPS model which
can be also reckoned as a variant of the SHS model with penetrable
core. Upon reversing the sign of the attractive energy scale
$\epsilon_a$ we obtain a PSS model with successive soft repulsive
barriers of decreasing height. If the second barrier is higher that
the first ($-\epsilon_a>\epsilon_r>0$), we find a  {potential that
is attractive for short distances and repulsive for larger
distances. An interesting  situation, that we have denoted as
HPS, corresponds to $-\epsilon_a>\epsilon_r=0$. In the limit of zero
temperature (or, equivalently, $\epsilon_a\to-\infty$) the HPS model
becomes the HHS model, characterized by an infinitely high barrier
 between $\sigma$ and $\sigma+\Delta$. Here
the equilibrium state consists of ``chains'' of connected particles:
two adjacent particles of the same chain move freely, provided that
the distance between their centers does not exceed $\sigma$; on the
other hand, particles of different chains behave as hard spheres of
diameter $\sigma+\Delta$.} In the limit $|\epsilon_a| \to 0$
 {(and also if $-\epsilon_a=\epsilon_r>0$)} the PSW fluid
reduces to the PS one, and all results obtained here are consistent
with previous analysis on the PS model within this limit. Finally,
all results smoothly converge to the HS limit when $\epsilon_r \to
\infty$ and $|\epsilon_a| \to 0$  {(or
$-\epsilon_a=\epsilon_r\to\infty$)}, as expected.

The combined effect of the absence of a hard core and the presence
of a finite attractive part of the PSW potential raises the
issue of the existence of a well-defined thermodynamic limit of the
system. We have analyzed this issue in detail and we have assessed
the limits of stability as a function of the ratio between the
attractive and repulsive energy scales: when $\epsilon_a/ \epsilon_r
\le \frac{1}{2}$ the system is stable whereas in the opposite case
the system might be unstable (when $\epsilon_r>0$ and $\epsilon_a<0$
the system is always stable independently of the energy scales). The
SPS limit turns out to be always unstable since the exact
fourth-order virial coefficient diverges, unlike the corresponding
SHS counterpart which is well-behaving.

A main advantage of exact relations is that one can asses the
reliability of approximate theories. A comparison with PY and HNC
closures unveils the corresponding strengths and weaknesses of both.
We have found that each of them has a domain in space where it
outperforms the other, and we have explained why this is so in terms
of the exact and approximate behavior of the second-order cavity
function $y_2(r)$. As a general feature, HNC tends to overestimate
the cavity function within the core whereas PY has the opposite
tendency. On the other hand, PY is consistently superior in the
large-$r$ region. Both approximate theories produce artifacts in the
low temperature region of the fourth virial coefficients.

It would be extremely interesting to extend the present work in some
respects.  While our analysis has provided a careful comparison of
the PY and HNC approximate theories with respect to the exact
result, we have not attempted a detailed physical
interpretation of the results. This is because our exact analysis
was limited to the lowest orders in density, which are
expected to be valid only within a rather limited region of the
phase diagram. It turns out, however, that even this limited
knowledge can be exploited to construct rather precise
approximations for the
 PSW model in the
limits of small penetrability  {($1-\gamma_r \ll 1$)} and high
penetrability  {($\gamma_r \ll 1$)}, for arbitrary value of the
density. This analysis mirrors that already performed for the PS
model \cite{Malijevsky06,Santos07}, can be tested against numerical
simulations, and nicely complements the exact low-density results
presented here. The boundedness of the class of penetrable-sphere
potentials raises the possibility of a phase transition even in a
one-dimensional system \cite{Acedo04,Cuesta04} and the presence of
the attractive part might also give rise to additional transitions
in the fluid phase. We plan to address this point in future work.

It is worth stressing that the large number of parameters present in
the PSW fluid (two energy scales, two characteristic lengths,
density, and temperature) may render the phase diagram analysis
quite problematic, so an exact understanding of the low-dimensional
behavior, where the analysis can be carried out almost fully
analytically, is always welcome. Having done this, the extension to
three dimensions should be facilitated, and our results predict an
extremely rich phase diagram which might be useful to describe
complex fluids with soft cores within a unified picture.

\appendix
\section{Ruelle's stability criterion}
\label{app:ruelle}
According to Ruelle's criterion, a sufficient condition of
thermodynamic stability is \cite{Fisher66,Ruelle69}
\begin{equation}
U_N(x_1,\ldots,x_N)=\sum_{i=1}^{N-1}\sum_{j=i+1}^N
\phi(|x_i-x_j|)\geq -NB
\end{equation}
for all configurations $\{x_i\}$, where $B$ is a fixed bound.

Let us first demonstrate that for small repulsion ($\epsilon_r < 2
\epsilon_a$) there exists at least one configuration violating the
stability constraint. We consider a particular configuration  where
the $N$ particles are distributed into $N/s$ overlapping clusters of
$s$ particles each, so that the centers of the $s$ particles
belonging to the same cluster coincide (with a tolerance
$\Delta/2$). Next, the centers of two adjacent clusters are
separated by a distance between $\sigma$ and $\sigma+\Delta$. The
potential energy corresponding to this configuration is
\begin{eqnarray}
U_N(s)&=&\frac{N}{s}\frac{s(s-1)}{2}\epsilon_r-
\left(\frac{N}{s}-1\right)s^2\epsilon_a~,
\label{ruelle:eq1}
\end{eqnarray}
 {The first term on the right-hand side represents the repulsive
energy of the $s(s-1)/2$ pairs of each cluster, times the number of
clusters. The second term is the attractive energy of the
interaction between the $s$ particles of each cluster and the $s$
particles of its neighbor cluster, times the number of pairs of
adjacent clusters. Obviously, the value of the total potential
energy $U_N$ depends on the value of $s$.} The extreme cases are
$s=1$ and $s=N$. We then see that the value that minimizes $U_N(s)$
is
\begin{eqnarray}
s_*&=&\frac{N}{2}\left(1-\frac{\epsilon_r}{2\epsilon_a}\right)~,
\label{ruelle:eq2}
\end{eqnarray}
which is meaningful only if $\epsilon_r<2\epsilon_a$.
The corresponding minimum value of $U_N(s)$ is
\begin{eqnarray}
U_N(s_*)&=&-N\left[\frac{\epsilon_r}{2}+\frac{N}{16}\epsilon_a
\left(2-\frac{\epsilon_r}{\epsilon_a}\right)^2\right]
~,~~~\epsilon_r<2\epsilon_a~.
\label{ruelle:eq3}
\end{eqnarray}
 {The quantity enclosed between brackets grows linearly with $N$
and so it is not bounded.} Therefore, if $\epsilon_r<2\epsilon_a$,
there exists at least a configuration that violates Ruelle's
criterion. On the other hand, we note that if
$\epsilon_r>2\epsilon_a$ the minimum of $U_N(s)$ is reached at
$s=1$, in which case
\begin{eqnarray}
U_N(s)\ge U_N(s=1) = -(N-1)\epsilon_a~,~~~\epsilon_r>2\epsilon_a~,
\label{ruelle:eq4}
\end{eqnarray}
so that all these special configurations are consistent with
Ruelle's criterion. Indeed we now show that no other configurations
violate Ruelle's criterion if $\epsilon_r>2\epsilon_a$ so that the
model is thermodynamically stable if the above condition is
satisfied.

Without loss of generality we can see any given configuration of $N$
particles as a set of $M$ clusters $(1\le M\le N)$,  {each
cluster $i$ being made of $s_i$   \emph{overlapping} particles
(i.e., any pair of particles of a given cluster are separated a
distance smaller than $\sigma$). For fixed $M$ and $\{s_i\}$, the
total potential energy can be decomposed as}
\begin{equation}
U_N(\{s_i\};M)=U_N^{\text{intra}}(\{s_i\};M)+U_N^{\text{inter}}(\{s_i\};M),
\end{equation}
where
\begin{eqnarray}
U_N^{\text{intra}}(\{s_i\};M)&=&\frac{\epsilon_r}{2}\sum_{i=1}^M
s_i(s_i-1)~
\label{ruelle:eq5}
\end{eqnarray}
 {is the contribution associated with pairs of particles inside
each cluster and $U_N^{\text{inter}}$ is the contribution associated
with  pairs of particles belonging in different clusters. Note that
in the latter contribution the energy for each pair can be
$\epsilon_r$ (if the separation is smaller than $\sigma$),
$-\epsilon_a$ (if the separation lies between $\sigma$ and
$\sigma+\Delta$), or zero (if the separation is larger than
$\sigma+\Delta$). It is clear that the minimum value of
$U_N^{\text{inter}}$} is achieved when all the particles of a
cluster interact attractively with all the particles of the neighbor
cluster:
\begin{eqnarray}
U_N^{\text{inter}}(\{s_i\};M)&\geq&-\epsilon_a\sum_{i=1}^{M-1}
s_is_{i+1}> -\epsilon_a\sum_{i=1}^{M-1} s_is_{i+1}-\epsilon_a
s_1s_M~.
\label{ruelle:eq6}
\end{eqnarray}
 {Therefore,}
\begin{eqnarray}
U_N(\{s_i\};M)&>&\frac{\epsilon_r}{2}\sum_{i=1}^M
s_i(s_i-1)-\frac{\epsilon_a}{2}\sum_{i=1}^{M}
s_i(s_{i-1}+s_{i+1})\equiv W_N(\{s_i\};M)~,
\label{ruelle:eq7}
\end{eqnarray}
where $s_0=s_M$ and $s_{M+1}=s_1$. Given $M$, what is the set of
population numbers $\{s_i\}$ that minimizes $W_N$ subject to
the constraint that $\sum_{i=1}^Ms_i=N$? Using a Lagrange multiplier
$\lambda$, the problem reduces to solve
\begin{eqnarray}
\frac{\partial}{\partial s_j}
\left[W_N(\{s_i\};M)-\lambda\sum_{i=1}^Ms_i\right]=
\frac{\epsilon_r}{2}(2s_j-1)-\epsilon_a(s_{j-1}+s_{j+1})-\lambda
=0~,~~~1\le j\le M~.
\label{ruelle:eq8}
\end{eqnarray}
 The solution is $s_i=N/M$ and
$\lambda=(N/M)(\epsilon_r-2\epsilon_a)-\epsilon_r/2$. This could
have been expected by symmetry arguments. Therefore, given $M$
clusters, the minimum $W_N$ is obtained with a uniform distribution
$s_i=s=N/M$. For $\epsilon_r>2\epsilon_a$ we can thus write
\begin{eqnarray} \nonumber
U_N(\{s_i\};M)&>&W_N(\{s_i=N/M\};M)=
\frac{\epsilon_r}{2}M(N/M)(N/M-1)-\epsilon_a M (N/M)^2\\
&=&(\epsilon_r/2-\epsilon_a)N^2/M-N\epsilon_r/2>
-N\epsilon_r/2~,
\label{ruelle:eq9}
\end{eqnarray}
which proves that Ruelle's stability criterion is satisfied.
\section{ {Second-order} cavity functions for SW}
\label{app:sw}
 {The first-order term $y_1^{(\SW)}(r)$ (for $\Delta < \sigma
\equiv 1$) is given by Eq.\ (\ref{exactpsw:eq2}) with $\gamma_r=1$.
This allows for a straightforward determination of
$y_{2C}^{(\SW)}(r)$ as}
\begin{eqnarray}
y_{2C}^{(\SW)}\left(r\right) &=&
\left[y_1^{(\SW)}\left(r\right)\right]^2.
\label{appsw:eq2}
\end{eqnarray}
Next, one can also evaluate the  {Fourier transform of the
integral corresponding to the $2A$ diagram. Going back to real
space, the result is}
\begin{eqnarray} \nonumber
y_{2A}^{(\SW)} \left(r\right) &=& -\frac{3}{2} \gamma
\left(1+\gamma\right)^2 \left(1-\Delta -r\right)^2
\Theta\left(1-\Delta -r\right) + \frac{3}{2} \left(1+\gamma\right)
\left(1+2 \gamma + 3 \gamma^2\right)\left(1-r\right)^2
\Theta\left(1-r\right)
\\ \nonumber
&&- \frac{3}{2} \gamma \left(2+4 \gamma + 3 \gamma^2\right) \left(1+\Delta -r\right)^2
\Theta\left(1+\Delta -r \right) + \frac{3}{2} \gamma^2 \left(1+\gamma\right)
\left(1+2 \Delta-r\right)^2 \Theta\left(1+2\Delta -r \right) \\ \nonumber
&&- \frac{1}{2} \left(1+\gamma\right)^3 \left(3-r\right)^2 \Theta\left(3-r\right)+
\frac{3}{2} \gamma \left(1+\gamma\right)^2 \left(3+\Delta-r\right)^2
\Theta\left(3+\Delta -r\right) \\
&& - \frac{3}{2} \gamma^2 \left(1+\gamma\right) \left(3 + 2\Delta
-r\right)^2 \Theta\left(3+2 \Delta -r \right) + \frac{1}{2} \gamma^3
\left(3 + 3 \Delta -r\right)^2 \Theta\left(3+3 \Delta -r\right).
\label{appsw:eq3}
\end{eqnarray}
For $y_{2B}^{(\SW)}(r)$ we can make use of the identity
\begin{eqnarray}
y_{2B}\left(r\right) &=& \int_{-\infty}^{\infty}
ds~y_1\left(s\right) f\left(s\right) f\left(\vert r-s\vert\right),
\label{appsw:eq4}
\end{eqnarray}
which leads to the result
\begin{eqnarray}
y_{2B}^{(\SW)}\left(r\right) &=& \nonumber \gamma
\left(1+\gamma\right) \left(2 -\Delta - 4 \gamma \Delta\right)
\left(\Delta -r \right) \Theta\left(\Delta -r\right)+ \gamma
\left(1+\gamma\right)^2 \left(1-\Delta -r\right)^2
\Theta\left(1-\Delta - r\right) \\ \nonumber && - \left(1+3 \gamma +
5 \gamma^2 + 3 \gamma^3\right)\left(1-r\right)^2
\Theta\left(1-r\right)+ \gamma \left(2 + 4 \gamma + 3
\gamma^2\right) \left(1+\Delta -r \right)^2\Theta\left(1+\Delta -r
\right) \\ \nonumber && -\gamma^2\left(1+\gamma\right)
\left(1+2\Delta-r\right)^2 \Theta\left(1+ 2 \Delta -r\right)+
\frac{1}{2} \left(1+\gamma\right)^2 \left(4-r-4 \gamma
\Delta\right)\left(2-r\right)\Theta\left(2-r\right) \\
&& - \gamma \left(1+\gamma\right)\left(4-r -4\gamma \Delta\right)
\left(2+\Delta -r \right) \Theta\left(2+ \Delta -r \right)+
\frac{1}{2} \gamma^2 \left(4-r-4 \gamma \Delta\right) \left(2+2
\Delta -r\right)\Theta\left(2+2 \Delta -r\right).\nonumber\\
\label{appsw:eq5}
\end{eqnarray}

Computation of  $y_{2D}^{(\SW)}(r)$ is much more laborious and
requires a different route. We go back to the general formalism and
compute the exact $\widetilde{\Omega}(s)$ from the Laplace transform
(\ref{formalism:eq1}), which is
\begin{eqnarray}
\widetilde{\Omega}\left(s\right) &=& \frac{e^{-s}}{s}
\left(1+\gamma- \gamma e^{-s \Delta} \right).
\label{appsw:eq6}
\end{eqnarray}
Equation (\ref{formalism:eq4}) then yields for the parameter $\xi$
the following density expansion:
\begin{eqnarray}
\xi &=& \rho + \left(1-\gamma \Delta\right)\rho^2 + \left[1-\gamma
\Delta \left(2-\Delta -2 \gamma \Delta \right) \right] \rho^3 +
\cdots .
\label{appsw:eq7}
\end{eqnarray}
Inserting this solution into Eq.\ (\ref{formalism:eq3})  and
inverting the Laplace transform (\ref{formalism:eq2}), we can obtain
the corresponding radial distribution function $g_{2}(r)$ correct up
to second order in density. Use of Eq.\ (\ref{exactpsw:eq1}) then
yields the corresponding cavity function $y_{2}(r)$ and then
$y_{2D}^{(\SW)}(r)$ is given by the difference
\begin{eqnarray}
y_{2D}^{(\SW)}\left(r\right)&=& 2 y_{2}\left(r\right)- 2
y_{2A}^{(\SW)}\left(r\right) -4 y_{2B}^{(\SW)}\left(r\right)-4
y_{2C}^{(\SW)}\left(r\right).
\label{appsw:eq8}
\end{eqnarray}
This provides the result for $r\ge 1$. Inside the core we have three
different regions under the assumption that $\Delta \le 1$, namely
$0 \le r \le \Delta$, $\Delta \le r \le 1-\Delta$ and $1-\Delta \le
r \le 1$. The quadratic expression in each region can be obtained by
imposing continuity conditions and with some help from numerical
evaluation. The final analytic result is
\begin{eqnarray}
y_{2D}^{(\SW)}\left(r\right) &=& \nonumber -2 \gamma
\left(1+\gamma\right) \left[ \left(1+\gamma+\gamma^2\right)r-2
+\Delta \left(1+3 \gamma -\gamma^2\right)\right] \left(\Delta -r
\right) \Theta\left(\Delta-r\right)
\\ \nonumber
&& - \gamma \left(1+\gamma\right)^2 \left(1-\Delta -r\right)^2 \Theta\left(1-\Delta -r\right)
+\left(1+3 \gamma + 5 \gamma^2+ 3 \gamma^3\right) \left(1-r\right)^2 \Theta\left(1-r\right)
\\ \nonumber
&& - \gamma \left(2+4 \gamma+3 \gamma^2\right) \left(1+\Delta-r\right)^2
\Theta\left(1+\Delta-r\right)+\gamma^2 \left(1+\gamma\right) \left(1+2\Delta -r\right)^2
\Theta\left(1+2 \Delta -r \right)
\\ \nonumber
&& + \left(1+\gamma\right)^2 \left[r\left(1-2\gamma-\gamma^2\right)-2+4\gamma+2\gamma^2+4 \gamma \Delta
\right] \left(2-r\right) \Theta\left(2-r\right)
\\
&& + 4 \gamma^2 \left(1+\gamma\right)\left(r-2-\Delta+\gamma \Delta\right)\left(2+\Delta-r\right)
\Theta\left(2+\Delta -r \right) -\gamma^4 \left(2+2 \Delta -r\right)^2
\Theta\left(2+2 \Delta -r\right).
\label{appsw:eq9}
\end{eqnarray}
Note that the first derivative $y^{\prime}(r)$ is
discontinuous at $r=\Delta$, $2$, $2+\Delta$, and $2+2\Delta$, as it
can be inferred from its explicit computation at this order in
density.
\section{ {Calculation of $B_4$ for the PSW model in the PY and HNC approximations}}
\label{app:B4}
 {Here the fourth virial coefficient predicted by the PY and HNC
approximations from the various thermodynamic routes
($v=\text{virial}$, $c=\text{compressibility}$, $e=\text{energy}$)
are given.}
\subsection{PY approximation}
 Using Eq.\ (\ref{PY_HNC:eq2}),
along with the recursion relations
(\ref{virial:eq12})--(\ref{virial:eq17}), we have
\begin{eqnarray}
B_4^{\text{PY},v}&=&\gamma_r^5\left[3-\gamma\Delta\left(9-7\Delta-16\gamma
\Delta+\Delta^2+6\gamma
\Delta^2+6\gamma^2\Delta^2-2\gamma^3\Delta^2\right) \right]
\nonumber \\&&-\frac{\gamma_r^4}{2} \left[4-\gamma
\Delta\left(12-6\Delta-18\gamma \Delta+ \Delta^2+3 \gamma
\Delta^2+3\gamma^2\Delta^2-3\gamma^3\Delta^2 \right)\right],
\label{app_PY:eq1}
\end{eqnarray}
\begin{eqnarray}
\chi_4^{\text{PY}}=-4\left(2B_2^3-3B_2B_3+B_4^{\text{PY},c}\right),
\label{app_PY:eq2}
\end{eqnarray}
\begin{eqnarray}
B_4^{\text{PY},c}&=&\frac{\gamma_r^5}{3}\left[7-\gamma\Delta\left(21-15\Delta-
36 \gamma \Delta+3\Delta^2+16 \gamma \Delta^2+16\gamma^2\Delta^2-4\gamma^3
\Delta^2\right)\right]
 \nonumber\\
&&-\frac{\gamma_r^4}{3} \left[4-\gamma
\Delta\left(12-6\Delta-18\gamma \Delta+ \Delta^2+3\gamma
\Delta^2+3\gamma^2\Delta^2-3\gamma^3\Delta^2\right)\right],
\label{app_PY:eq3}
\end{eqnarray}
\begin{eqnarray}
u_4^{\text{PY}}&=& \frac{\gamma_r^3}{6}\left(1+\gamma_r \gamma
\right) \Delta \left[12-18 \gamma_r -6 \Delta
\left(1-2\gamma_r+6\gamma- 10 \gamma_r \gamma \right)
+\Delta^2\left(1-2\gamma_r+6\gamma-26 \gamma_r \gamma+9
\gamma^2-36\gamma_r \gamma^2\right.\right. \nonumber\\
&&\left.\left.-12\gamma^3+16 \gamma_r \gamma^3\right)\right]
\epsilon_a -\frac{\gamma_r^3}{6}\left(1-\gamma_r \right)
\Delta\left[16-28\gamma_r-6\gamma \Delta\left(6-11\gamma_r \right)
+6\gamma \Delta^2\left(3-8\gamma_r+6\gamma-14\gamma_r \gamma
\right)\right.
 \nonumber \\
&& \left.-\gamma\left(1+\gamma \right)\Delta^3
\left(3-10\gamma_r+3\gamma-28\gamma_r \gamma \right)
\right]\epsilon_r.
\label{app_PY:eq4}
\end{eqnarray}
The fourth virial coefficient associated with the energy route,
$B_4^{\text{PY},e}$, is obtained from  {Eq.\ \eqref{app_PY:eq4}
as}
\begin{eqnarray}
B_4^{\text{PY},e}=3\int_0^\beta d\beta'\, u_4^{\text{PY}}(\beta').
\label{app_PY:eq5}
\end{eqnarray}
Its expression  is quite long and so it is omitted here. In addition
to its dependence on $\gamma_r$ and $\gamma$, $B_4^{\text{PY},e}$
depends on $\epsilon_a/(n_a\epsilon_a-n_r\epsilon_r)$ with
$(n_a,n_r)=(1,1),(1,2),(1,3),(1,4),(2,1),(2,3),(3,1),(3,2),(4,1)$.

It is instructive to consider some special cases. First, the results for the PS
model correspond to the limit $\epsilon_a\to 0$ ($\gamma \to 0$) or
$\epsilon_a\to -\epsilon_r$ ($\gamma \to -1$):
\begin{eqnarray}
\lim_{\epsilon_a\to 0} B_4^{\text{PY},v}=\lim_{\epsilon_a\to
-\epsilon_r}\frac{B_4^{\text{PY},v}}{(1+\Delta)^3}=\gamma_r^4(3\gamma_r-2),
\label{app_PY:eq6}
\end{eqnarray}
\begin{eqnarray}
\lim_{\epsilon_a\to 0} B_4^{\text{PY},c}=\lim_{\epsilon_a\to
-\epsilon_r}\frac{B_4^{\text{PY},c}}{(1+\Delta)^3}=
\gamma_r^4\left(\frac{7\gamma_r}{3}-\frac{4}{3}\right),
\label{app_PY:eq7}
\end{eqnarray}
\begin{eqnarray}
\lim_{\epsilon_a\to 0}B_4^{\text{PY},e}=\lim_{\epsilon_a\to
-\epsilon_r}\frac{B_4^{PY,e}}{(1+\Delta)^3}=\gamma_r^4
\left(\frac{14 \gamma_r}{5}-2\right).
\label{app_PY:eq8}
\end{eqnarray}
 In the special case of the HPS model ($\epsilon_a<0$
and $\epsilon_r\to 0$) one finds that $B_4^{\text{PY},v}$ and
$B_4^{\text{PY},e}$ reduce to the exact result  [see Eq.\
\eqref{app_exact:eq6}] but $\lim_{\epsilon_r\to
0}B_4^{\text{PY},c}=-\gamma_a^4\Delta^3$.

 The conventional SW  model corresponds to
$\epsilon_r\to\infty$ ($\gamma_r \to 1$):
\begin{eqnarray}
\lim_{\epsilon_r\to \infty}B_4^{\text{PY},v}=1-\gamma \Delta
\left(3-4\Delta-7\gamma \Delta+\frac{1}{2}\Delta^2+\frac{9}{2} \gamma \Delta^2
+\frac{9}{2}\gamma ^2\Delta^2-\frac{1}{2}\gamma^3\Delta^2\right),
\label{app_PY:eq9}
\end{eqnarray}
\begin{eqnarray}
\lim_{\epsilon_r\to \infty}B_4^{\text{PY},c}=1-\gamma \Delta\left(3-
3\Delta-6\gamma \Delta+\frac{2}{3}\Delta^2+\frac{13}{3}\gamma \Delta^2
+\frac{13}{3}\gamma^2\Delta^2-\frac{1}{3}\gamma^3\Delta^2\right),
\label{app_PY:eq10}
\end{eqnarray}
\begin{eqnarray}
\lim_{\epsilon_r\to \infty}B_4^{\text{PY},e}=\frac{4}{5}-\gamma
\Delta\left(3-3\Delta-6\gamma \Delta+\frac{1}{2}\Delta^2+5\gamma \Delta^2
+\frac{9}{2} \gamma^2\Delta^2-\frac{1}{2}\gamma^3\Delta^2\right).
\label{app_PY:eq11}
\end{eqnarray}
 {If, furthermore, the SHS limit ($\gamma\to\infty$ and
$\Delta\to 0$ with $\gamma\Delta=\text{const}$) is taken in Eqs.\
(\ref{app_PY:eq9})--(\ref{app_PY:eq11}), an artificial divergence of
$B_4$ is obtained.}

The  results corresponding to HS  are obtained by taking either the
limit $\epsilon_r\to\infty$ ($\gamma_r\to 1$) in Eqs.\
(\ref{app_PY:eq6})--(\ref{app_PY:eq8}) or the limit $\epsilon_a\to
0$ ($\gamma \to 0$) in Eqs.\
(\ref{app_PY:eq9})--(\ref{app_PY:eq11}). In either case one
sees that the virial and compressibility routes yield the exact
result, while the energy route value is wrong by a factor $4/5$. A
third possibility consists of taking the limit
$\epsilon_a\to-\infty$ ($\gamma \to -1$) in Eqs.\
(\ref{app_PY:eq9})--(\ref{app_PY:eq11}). However,  in this latter
case the energy route yields an incorrect dependence on $\Delta$:
\begin{eqnarray}
\lim_{\epsilon_a\to-\infty}\lim_{\epsilon_r\to
\infty}B_4^{\text{PY},e}=\frac{4}{5}+
\frac{\Delta}{2}\left(6+6\Delta+\Delta^2\right).
\label{app_PY:eq12}
\end{eqnarray}
The  fact that the right-hand side of Eq.\ (\ref{app_PY:eq12}) is
not proportional to $(1+\Delta)^3$ implies that if one starts from
$B_4^{\text{PY},e}$ for the PSS model of shoulder height and width
$-\epsilon_a$ and $\Delta$, respectively, and then one takes the
limit $\epsilon_a\to-\infty$ to get the HS model of diameter
$1+\Delta$, the result has an artificial dependence on $\Delta$.
This anomaly of the PY description was discussed in Ref.\
\cite{Santos06}.

\subsection{HNC approximation}
Similarly to the preceding analysis, from Eq.\
(\ref{PY_HNC:eq1}) one gets
\begin{eqnarray}
B_4^{\text{HNC},v}=\frac{3}{2}B_4^{\text{PY},c},
\label{app_HNC:eq1}
\end{eqnarray}
\begin{eqnarray}
\chi_4^{\text{HNC}}=-4\left(2B_2^3-3B_2B_3+B_4^{\text{HNC},c}\right),
\label{app_HNC:eq2}
\end{eqnarray}
\begin{eqnarray}
B_4^{\text{HNC},c}&=&\frac{5\gamma_r^5}{12}\left[7-\gamma \Delta
\left(21-15\Delta-36\gamma
\Delta+3\Delta^2+16\gamma \Delta^2+16\gamma^2\Delta^2-4\gamma^3\Delta^2\right)\right]
 \nonumber \\
&&-\frac{\gamma_r^4}{2}
\left[4-\gamma\Delta\left(12-6\Delta-18\gamma\Delta+\Delta^2+
3\gamma\Delta^2+3\gamma^2\Delta^2-3\gamma^3\Delta^2\right)\right],
\label{app_HNC:eq3}
\end{eqnarray}
\begin{eqnarray}
u_4^{\text{HNC}}=\frac{1}{3}\frac{\partial}{\partial\beta}B_4^{\text{HNC},v},
\label{app_HNC:eq4}
\end{eqnarray}
Equation (\ref{app_HNC:eq4})  {implies that
$B_4^{\text{HNC},e}=B_4^{\text{HNC},v}$. This} confirms that, in
general, the energy and virial routes are thermodynamically
consistent in the HNC approximation \cite{Morita60}. It is also
noteworthy that  the fourth virial coefficient predicted by the HNC
approximation in the virial and energy routes is exactly three
halves the one predicted by the PY approximation in the
compressibility energy route, Eq.(\ref{app_HNC:eq1}). This simple
relation is not restricted to 1D models since it also occurs in the
3D PS model \cite{Santos07}. It would be extremely interesting to
check whether relation (\ref{app_HNC:eq1}) is a general property
valid for any interaction potential and for any dimensionality.

In the PS and SW  limits Eq.\ (\ref{app_HNC:eq3}) becomes
\begin{eqnarray}
\lim_{\epsilon_a\to 0}B_4^{\text{HNC},c}=\lim_{\epsilon_a\to
-\epsilon_r}\frac{B_4^{\text{HNC},c}}{(1+\Delta)^3}=\gamma_r^4
\left(\frac{35\gamma_r}{12}-2\right),
\label{app_HNC:eq6}
\end{eqnarray}
\begin{eqnarray}
\lim_{\epsilon_r\to \infty}B_4^{\text{HNC},c}=\frac{11}{12}-\gamma
\Delta\left(\frac{11}{4}- \frac{13}{4}\Delta-6\gamma
\Delta+\frac{3}{4}\Delta^2+ \frac{31}{6} \gamma \Delta^2
+\frac{31}{6}\gamma^2\Delta^2-\frac{1}{6}\gamma^3\Delta^2\right),
\label{app_HNC:eq7}
\end{eqnarray}
respectively.

 The three routes in the HNC theory yield the exact result
\eqref{app_exact:eq6} in the HPS limit. However, as in the
case of the PY theory, an artificial divergence of $B_4$ is
predicted in the SHS limit.

\begin{acknowledgments}
 We are grateful to C. Likos for suggesting us to look at the
problem of Ruelle's instability. The research of A.S. has been
supported by  the Ministerio de Educaci\'on y Ciencia (Spain)
through Grant No.\ FIS2007--60977. A.G. and R.F. acknowledge support
from the Italian MIUR (PRIN-COFIN 2006/2007).
\end{acknowledgments}


\begin{thebibliography}{99}

\bibitem{Barrat03}
J.L. Barrat and J.P. Hansen \textit{Basic Concepts for Simple and
Complex Liquids} (Cambridge University Press, Cambridge, 2003).

\bibitem{Likos01}
 C.N. Likos, Phys. Rep. \textbf{348}, 267 (2001).

\bibitem{Marquest89}
 C. Marquest and T.A. Witten, J. Phys. (France) \textbf{50},
1267 (1989).

\bibitem{Malijevsky06}
 Al. Malijevsk\'y and A. Santos, J. Chem. Phys.
\textbf{124}, 074508 (2006).

\bibitem{Salsburg53}
 Z.W. Salsburg, R.W. Zwanzig, and J.G. Kirkwood, J. Chem.
Phys. \textbf{21}, 1098 (1953).

\bibitem{Tonks36}
 E. Tonks, Phys. Rev. \textbf{50}, 955 (1936).

\bibitem{Feynman72}
 R.P. Feynman \textit{Statistical Mechanics}
(W.A. Benjamin, Reading, 1972).

\bibitem{Corti98}
 D.S. Corti and P.G. Debenedetti, Phys. Rev. E \textbf{57},
4211 (1998).

\bibitem{Lebowitz71}
 {J.L. Lebowitz and D. Zomick, J. Chem. Phys. \textbf{54}, 3335
(1971).}

\bibitem{Santos07a}
 {A. Santos,  Phys. Rev. E \textbf{76}, 062201 (2007).}

\bibitem{Seaton86}
 N.A. Seaton and E.D. Glandt, J. Chem. Phys. \textbf{84},
4595 (1986).

\bibitem{Yuste93}
 S.B. Yuste and A. Santos, J. Stat. Phys. \textbf{72},
703 (1993).

\bibitem{Heying04}
M. Heying and D.S. Corti, Fluid Phase Equilibria \textbf{220}, 85
(2004).

\bibitem{Lieb66}
 E.H. Lieb and D.C. Mattis, \textit{Mathematical physics
in one dimension} (Academic Press, New York and London, 1966);
 {D. C. Mattis, \emph{The Many-Body Problem: An Encyclopedia of
Exactly Solved Models in One Dimension} (World Scientific,
Singapore, 1993).}

\bibitem{Fantoni03}
 R. Fantoni, Ph.D. thesis, University of Trieste, 2003
(unpublished).

\bibitem{note1}
 It may be useful to stress that in quantum
statistical mechanics the condition of impenetrability in a
one-dimensional fluid with periodic boundary conditions is
responsible for the impossibility to treat the particles just as
fermions or bosons but one necessarily has to introduce the anyonic
fractional statistics. In classical statistical mechanics the change
in the topology of phase space does not have such a dramatic
consequence.

\bibitem{Santos07}
A. Santos and Al. Malijevsk\'y,  Phys. Rev. E \textbf{75}, 021201
(2007).

\bibitem{Fisher66}
 M.E. Fisher and D. Ruelle, J. Math. Phys. \textbf{7},
260 (1966).

\bibitem{Ruelle69}
 D. Ruelle, \textit{Statistical Mechanics: Rigorous Results}
(Benjamin, London, 1969).

\bibitem{Widom70} B. Widom and J.S. Rowlinson, J. Chem. Phys. \textbf{52},
1670 (1970).

\bibitem{Torquato84} S. Torquato, J. Chem. Phys.\textbf{81}, 5079 (1984)

\bibitem{Rikvold85} P.A. Rikvold and G. Stell, J. Collid Interface Sci., 
\textbf{108}, 158 (1985); P.A. Rikvold and G. Stell, J. Chem. Phys. 
\textbf{82}, 1014 (1985).

\bibitem{Louis00}
 A.A. Louis, P.G. Bolhuis, and J.P. Hansen,
\textbf{62}, 7961 (2000).

\bibitem{Heyes07}
 D.M. Heyes, M.J. Cass, and G. Rickayzen, J. Chem. Phys.
\textbf{126}, 084510 (2007).

\bibitem{Baxter68}
 R.J. Baxter, J. Chem. Phys. \textbf{49}, 2770 (1968).

\bibitem{Fantoni07}
 R. Fantoni, D. Gazzillo, A. Giacometti, M. Miller, and G.
Pastore, J. Chem. Phys. \textbf{127}, 234507 (2007).

\bibitem{Stell91}
 G. Stell, J. Stat. Phys. \textbf{63}, 1203 (1991).

\bibitem{Kikuchi55}
R. Kikuchi, J. Chem. Phys. \textbf{23}, 2327 (1955).

\bibitem{Elkoshi85}
 Z. Elkoshi, H. Reiss, and A.D. Hammerich, J. Stat. Phys.
\textbf{41}, 685 (1985).

\bibitem{Hansen86}
 J-P. Hansen and I.R. McDonald, \textit{Theory of Simple Liquids},  {3rd. ed.
(Academic Press, London, 2006).}

\bibitem{note2}
 {Note that the cancellation of these three singularities do not
occur in the HNC approximation, where $y_{2D}^{(\SHS)}(r)$ is
neglected. In the PY approximation both $y_{2C}^{(\SHS)}(r)$ and
$y_{2D}^{(\SHS)}(r)$ are neglected, so that the singularities (i)
and (ii) persist.}

\bibitem{Acedo04}
 L. Acedo and A. Santos, Phys. Lett. A \textbf{323},
427 (2004).

\bibitem{Cuesta04}
 J.A. Cuesta and A. S\'anchez, J. Stat. Phys.
\textbf{115}, 869 (2004).

\bibitem{Santos06}
 A. Santos,  Mol. Phys. \textbf{104}, 3411 (2006).

\bibitem{Morita60}
T. Morita, Prog. Theor. Phys. \textbf{23}, 829 (1960); see also p.\
636 of J. A. Barker and D. Henderson, Rev. Mod. Phys. \textbf{48},
587 (1976).


\end{thebibliography}

\end{document}